\documentclass[twocolumn,twoside,superscriptaddress,prl,aps]{revtex4}
\usepackage{amsmath,mathrsfs,amsbsy,color,graphicx,bm,amsthm,amsfonts,dsfont}
\usepackage{times}
\usepackage{units}
\usepackage{braket}
\usepackage{pdfpages}
\usepackage{standalone}

\begin{document}

\title{Secret sharing of a quantum state}

\author{He Lu}
\affiliation{Shanghai Branch, National Laboratory for Physical Sciences at Microscale and Department of Modern Physics, University of Science and Technology of China, Shanghai 201315, China}
\affiliation{Synergetic Innovation Center of Quantum Information and Quantum Physics, University of Science and Technology of China, Hefei, Anhui 230026, China}
\author{Zhen Zhang}
\affiliation{Center for Quantum Information, Institute for Interdisciplinary Information Sciences, Tsinghua University, Beijing 100084, China}

\author{Luo-Kan Chen}
\affiliation{Shanghai Branch, National Laboratory for Physical Sciences at Microscale and Department of Modern Physics, University of Science and Technology of China, Shanghai 201315, China}
\affiliation{Synergetic Innovation Center of Quantum Information and Quantum Physics, University of Science and Technology of China, Hefei, Anhui 230026, China}
\author{Zheng-Da Li}
\affiliation{Shanghai Branch, National Laboratory for Physical Sciences at Microscale and Department of Modern Physics, University of Science and Technology of China, Shanghai 201315, China}
\affiliation{Synergetic Innovation Center of Quantum Information and Quantum Physics, University of Science and Technology of China, Hefei, Anhui 230026, China}
\author{Chang Liu}
\affiliation{Shanghai Branch, National Laboratory for Physical Sciences at Microscale and Department of Modern Physics, University of Science and Technology of China, Shanghai 201315, China}
\affiliation{Synergetic Innovation Center of Quantum Information and Quantum Physics, University of Science and Technology of China, Hefei, Anhui 230026, China}
\author{Li Li}
\affiliation{Shanghai Branch, National Laboratory for Physical Sciences at Microscale and Department of Modern Physics, University of Science and Technology of China, Shanghai 201315, China}
\affiliation{Synergetic Innovation Center of Quantum Information and Quantum Physics, University of Science and Technology of China, Hefei, Anhui 230026, China}
\author{Nai-Le Liu}
\affiliation{Shanghai Branch, National Laboratory for Physical Sciences at Microscale and Department of Modern Physics, University of Science and Technology of China, Shanghai 201315, China}
\affiliation{Synergetic Innovation Center of Quantum Information and Quantum Physics, University of Science and Technology of China, Hefei, Anhui 230026, China}
\author{Xiongfeng Ma}
\affiliation{Center for Quantum Information, Institute for Interdisciplinary Information Sciences, Tsinghua University, Beijing 100084, China}

\author{Yu-Ao Chen}
\affiliation{Shanghai Branch, National Laboratory for Physical Sciences at Microscale and Department of Modern Physics, University of Science and Technology of China, Shanghai 201315, China}
\affiliation{Synergetic Innovation Center of Quantum Information and Quantum Physics, University of Science and Technology of China, Hefei, Anhui 230026, China}
\author{Jian-Wei Pan}
\affiliation{Shanghai Branch, National Laboratory for Physical Sciences at Microscale and Department of Modern Physics, University of Science and Technology of China, Shanghai 201315, China}
\affiliation{Synergetic Innovation Center of Quantum Information and Quantum Physics, University of Science and Technology of China, Hefei, Anhui 230026, China}

\date{\today}

\begin{abstract}
Secret sharing of a quantum state, or quantum secret sharing, in which a dealer wants to share certain amount of quantum information with a few players, has wide applications in quantum information.
The critical criterion in a threshold secret sharing scheme is \emph{confidentiality}; with less than the designated number of players, no information can be recovered. Furthermore, in a quantum scenario, one additional critical criterion exists; the capability of sharing entangled and unknown quantum information. Here by employing a six-photon entangled state, we demonstrate a quantum threshold scheme, where the shared quantum secrecy can be efficiently reconstructed with a state fidelity as high as 93$\%$. By observing that any one or two parties cannot recover the secrecy, we show that our scheme meets the confidentiality criterion. Meanwhile, we also demonstrate that entangled quantum information can be shared and recovered via our setting, which demonstrates that our implemented scheme is fully quantum. Moreover, our experimental setup can be treated as a decoding circuit of the 5-qubit quantum error-correcting code with two erasure errors.

\end{abstract}

\maketitle

Suppose two presidents have established a secure quantum channel via sharing of entangled states, such as many Einstein-Podosky-Rosen (EPR) pairs. At some point, one president takes a vacation and does not trust her individual vice presidents entirely; she therefore decides to divide up her halves of the EPR pairs into shares and distributes to the vice presidents in a quantum secret sharing (QSS) scheme. Only when all of the vice presidents work together are they allowed to communicate with the other president. Hence, in quantum cryptography, for instance, QSS can help to establish a quantum key in a multipartite scenario. Moreover, in a long-distance quantum network, the quantum channels, by which a quantum state can be transmitted between remote nodes, are typically very lossy. QSS is an efficient error correction protocol against qubit losses as erasure errors. Furthermore, QSS provides a robust and secure solution for quantum state storage and computation\cite{Cleve99}.

A significant class of secret sharing schemes is the $(k,n)$ threshold scheme, which is described as follows. The dealer encodes the initial secret into a large system composed of $n$ parts, and sends each player a share. To recover the dealer's information, at least $k$ (with $k\le n$) players should combine their shares together. Any subgroup with less than $k$ players is forbidden to decode any knowledge about the shared information. There are two criteria in a $(k,n)$ threshold scheme. The first criterion is \emph{reliability}; if more than $k$ players combine their shared pieces together, the information originated from the dealer can be faithfully recovered. The second criterion is \emph{confidentiality}\cite{Shamir1979,blakley1899safeguarding}; otherwise, with less than $k$ players, no information can be recovered. The no-cloning theorem\cite{wooters1982quantum,dieks1982communication} implies that no quantum $(k, n)$ threshold scheme exists for $2k\le n$. In QSS, a third critical criterion exists, namely the capability of sharing entangled and unknown quantum information, as required in the aforementioned quantum cryptography example. Numerous attempts  to realise QSS have been reported in literature; however, none of them satisfy all three criteria. For instance, in many experiments, quantum means are employed to share classical information\cite{tittel2001experimental,chen2005experimental,schmid2005experimental,gaertner2007experimental}, none of which can satisfy the confidentiality criterion when used for sharing a quantum state. In other implemented schemes\cite{PhysRevLett.92.177903,PhysRevA.71.033814,bell2014experimental}, pure-qubit state sharing has been demonstrated; however,  entangled states have never been shared and recovered.

In their seminal work, Cleve, Gottesman and Lo\cite{Cleve99} showed that for any $k\le n\le 2k-1$, efficient constructions of quantum threshold schemes exist. The essential idea is that a quantum $(k, 2k-1)$ threshold scheme can be realised by a quantum error-correcting code that is capable of correcting $k-1$ erasure errors with a code length of $2k-1$. Intuitively, if an error-correcting code can correct $k-1$ erasure errors, any $k$ shares can recover the initial state by treating the missing $k-1$ shares as erasure errors. Any information gain of the unknown initial state by measuring $k-1$ shares leads to disturbance of the recovered state by the remaining $k$ shares. Because $k$ shares can be used to perfectly reconstruct the initial state, no information can be obtained by less than $k-1$ shares, otherwise the principle of `information gain means disturbance' in quantum mechanics is violated\cite{Bennett:BBM92:1992}. Furthermore, for a general case with $n\le 2k-1$, the quantum $(k, n)$ threshold scheme can be construct by discarding $2k-1-n$ shares from a quantum $(k, 2k-1)$ threshold scheme.

According to the Cleve-Gottesman-Lo secret sharing theory\cite{Cleve99}, the $(3,3)$ threshold scheme is inherited from the 5-qubit quantum error-correcting code\cite{Laflamme96}. The derivations are shown in Appendix. The dealer holds the to-be-shared quantum state $\alpha\ket{H}+\beta\ket{V}$, which, in principle, could be unknown to the dealer, and encodes it into a three-photon mixed state
\begin{equation}\label{QSS:4pure}
\rho_{\text{QSS}}=\frac{1}{4}\sum\limits_{i,j=0}^{1}\ket{\phi_{ij}}\bra{\phi_{ij}}
\end{equation}
with
%\begin{equation} \label{QSS:rhoQSS}
%\begin{aligned}
%\rho_{\text{QSS}}=\frac{1}{4}\sum_{i,j}\ket{\phi_{i,j}}\bra{\phi_{i,j}},
%\end{aligned}
%\end{equation}
\begin{widetext}
\begin{equation} \label{QSS:4pure-1}
\begin{aligned}
\ket{\phi_{00}} &= \frac{1}{\sqrt{2}}(\alpha\ket{H}+\beta\ket{V})_A(\ket{HH}-\ket{VV})_{BC}-\frac{1}{\sqrt{2}}(\beta\ket{H}-\alpha\ket{V})_A(\ket{HH}+\ket{VV})_{BC}\\
\ket{\phi_{01}} &=\frac{1}{\sqrt{2}}(\alpha\ket{H}+\beta\ket{V})_A(\ket{HH}-\ket{VV})_{BC}+\frac{1}{\sqrt{2}}(\beta\ket{H}-\alpha\ket{V})_A(\ket{HH}+\ket{VV})_{BC}\\
\ket{\phi_{10}} &= \frac{1}{\sqrt{2}}(\alpha\ket{V}+\beta\ket{H})_A(\ket{HV}-\ket{VH})_{BC}-\frac{1}{\sqrt{2}}(\alpha\ket{H}-\beta\ket{V})_A(\ket{VH}+\ket{HV})_{BC}\\
\ket{\phi_{11}} &=\frac{1}{\sqrt{2}}(\alpha\ket{V}+\beta\ket{H})_A(\ket{HV}-\ket{VH})_{BC}+ \frac{1}{\sqrt{2}}(\alpha\ket{H}-\beta\ket{V})_{A}(\ket{VH}+\ket{HV})_{BC},\\
\end{aligned}
\end{equation}
\end{widetext}
where $\ket{H}$ and $\ket{V}$ denote horizontal and vertical polarisation, respectively. The subscripts $A$, $B$ and $C$ represent the three players Alice, Bob and Charlie in the scheme. Note that the to-be-shared quantum state can be unknown to the dealer. In fact, it can be a part of a large entangled state.

The quantum state of Eq.~\eqref{QSS:4pure} can also be treated as the 5-qubit code state after two erasure errors. Thus, if we can show that the original qubit can be recovered from Eq.~\eqref{QSS:4pure}, we can conclude that the 5-qubit code is capable of correcting two erasure errors, which has been proven to be equivalent to correcting an arbitrary error\cite{Grassl97}. Hence such s decoding circuit could also demonstrate that the 5-qubit code is capable of correcting an arbitrary qubit error.

\begin{figure*}[htbp]
\includegraphics[width=\linewidth]{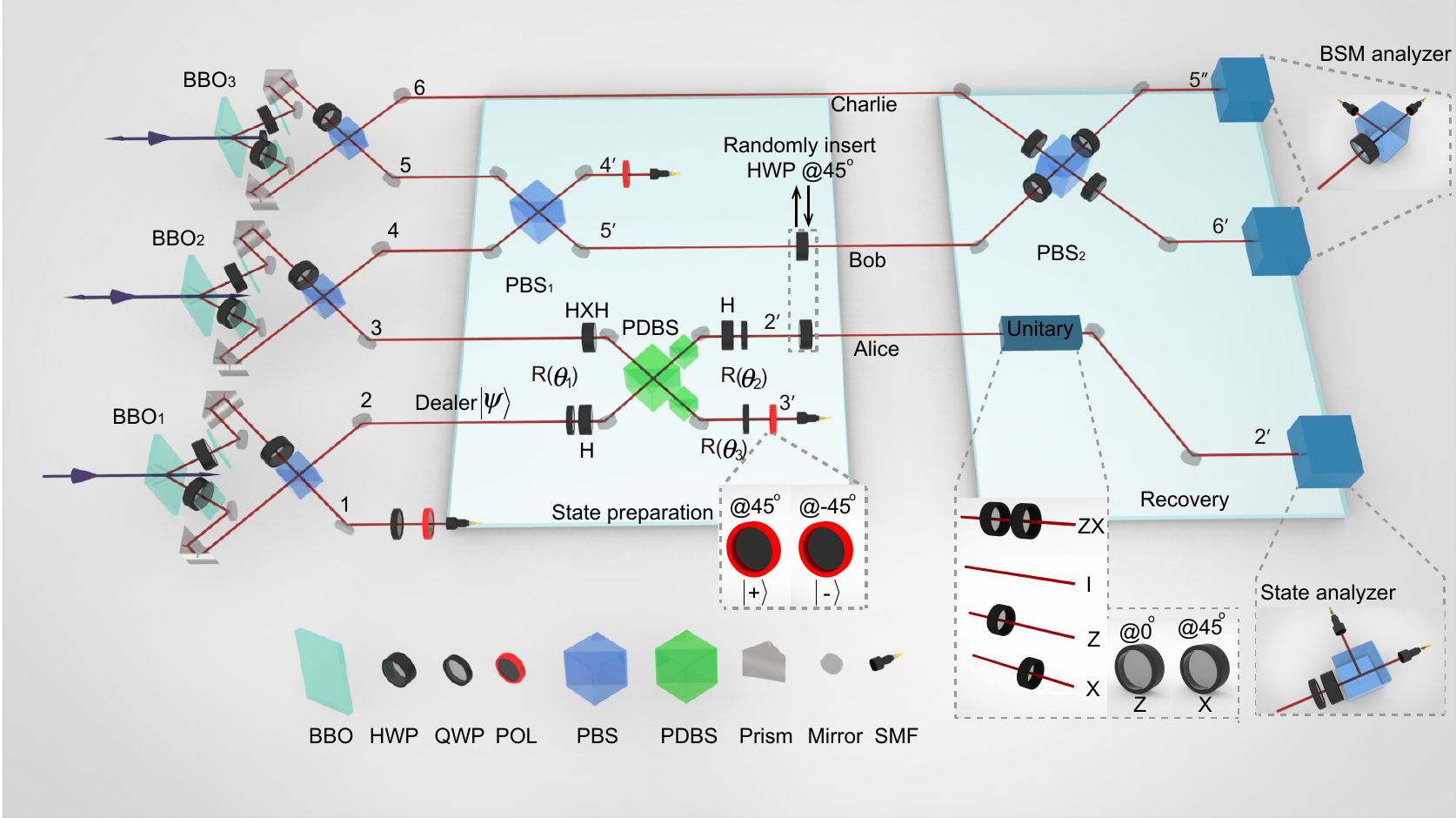}
\caption{Illustration of the experimental setup.
An ultraviolet (UV) pulse passes through BBO$_{1}$, BBO$_{2}$ and BBO$_{3}$ successively. After pumping on BBO$_{1}$ (BBO$_{2}$), the UV pulse is refocused by lenses and directed to BBO$_{2}$ (BBO$_{3}$) by mirrors (not shown here). The interference on PBS$_1$, PDBS and PBS$_2$ is obtained by finely adjusting the path length of the two input photons. To achieve good visibility of interference, we filter the photons temporally by narrow band filters and spatially by single-mode fibres. We observe the sixfold coincidence of 75 counts per hour (cps)  of \emph{confidentiality} testing and 14cph of \emph{reliability} testing (see Methods for more details).
}
\label{fig:setup}
\end{figure*}

To generate the three-photon mixed state $\rho_{\text{QSS}}$, we employ a six-photon entangled state. A schematic of the experimental setup is shown in Fig.~\ref{fig:setup}. An ultraviolet laser pulse ($\sim$ 140 fs, 76 MHz, 390 nm) successively passes through three 2-mm-thick $\beta$-barium borate (BBO) crystals to generate three entangled photon pairs\cite{Kwiat95}. We then overlap the two generated photons on a polasized beam splitter (PBS) to generate an ultra-bright entanglement photon pair\cite{Yao12}. For paths $i,j$, the entangled pair state is $\ket{\Phi^+}_{ij}=(\ket{HH}_{ij}+\ket{VV}_{ij})/\sqrt{2}$. Details of the photon source are presented in the Methods section. Using three entangled pairs, we prepare the state shown in Eq.~\eqref{QSS:4pure} for the $(3, 3)$ threshold scheme. In Fig.~\ref{fig:setup}, $X$, $Y$ and $Z$ are denoted as the Pauli operators and $H$ is denoted as the Hadamard operation.

The dealer (photon 2) holds the quantum secret $\alpha\ket{H}_{2}+\beta\ket{V}_{2}$ heralded by projecting photon 1 on the state $\alpha^{\ast}\ket{H}+\beta^{\ast}\ket{V}$. Photons 4 and 5 are overlapping on PBS$_1$, which leads the outgoing state a four-photon Greenberger-Horne-Zeilinger (GHZ) state $\ket{\text{GHZ}}_{4}=(\ket{HHHH}_{34^{\prime}5^{\prime}6}+\ket{VVVV}_{34^{\prime}5^{\prime}6})/\sqrt{2}$.
By projecting photon $4^{\prime}$ on the state $\ket{+}=(\ket{H}+\ket{V})/\sqrt{2}$, $\ket{\text{GHZ}}_{3}=(\ket{HHH}_{35^{\prime}6}+\ket{VVV}_{34^{\prime}5^{\prime}})/\sqrt{2}$ is obtained. A rotation $XHX$, which is realised by a half-wave plate (HWP) set at 67.5$^{\circ}$, is applied on photon 3 to convert the state $\ket{\text{GHZ}}_{3}$ into $(\ket{-HH}_{34^{\prime}5^{\prime}}-\ket{+VV}_{34^{\prime}5^{\prime}})/\sqrt{2}$, where $\ket{+}=(\ket{H}+\ket{V})/\sqrt{2}$ and $\ket{-}=(\ket{H}-\ket{V})/\sqrt{2} $. A controlled-$XZ$ gate, which can be decomposed into three phase shift gates (i.e., two Hadamard gates and a controlled-$Z$ (C-phase) gate; the decomposition of controlled-$XZ$ gate is shown in Appendix), is applied on target photon 2 and control photon 3.  The phase shift gate $R(\theta)$ keeps $\ket{H}$ unchanged but adds a phase $e^{i\theta}$ on $\ket{V}$, i. e.  $R(\theta)\ket{H}=\ket{H}$, $R(\theta)\ket{V}=e^{i\theta}\ket{V}$. The sequence of the three phase shift gate is shown in Fig.~\ref{fig:setup}, and the value of three phase shifts in our experiment are set to $\theta_1=-\pi/2$ and $\theta_2=\theta_3=\pi/2$. $R(\pi/2)$ and $R(-\pi/2)$ are achieve by setting the quarter-wave plates (QWPs) at $0^\circ$ and $90^\circ$. Two Hadamard gates (HWPs set at 22.5$^{\circ}$) are applied on target photon~2 and 2$^\prime$ before and after the C-phase gate. The C-phase gate is implemented by overlapping photons~1 and 3 on a polarisation-dependent beam splitter (PDBS) ($T_H=1$ and $T_V=1/3$) with two supplemental PDBSs ($T_V=1$ and $T_H=1/3$) at each exit port of the overlapping PDBS\cite{Kiesel05}. After the controlled-$XZ$ gate, the state becomes
\begin{widetext}
\begin{equation}
\begin{split}
&\frac{1}{\sqrt{2}}(\ket{-HH}_{35^{\prime}6}-\ket{+VV}_{34^{\prime}5^{\prime}})\otimes(\alpha\ket{H_{2}}+\beta\ket{V}_{2})\\
&\xrightarrow{C-XZ(3-2)} \frac{1}{\sqrt{2}}\ket{H}_{3^{\prime}} (\ket{HH}_{5^{\prime}6}-\ket{VV}_{5^{\prime}6})(\alpha\ket{H}_{2^{\prime}}+\beta\ket{V}_{2^{\prime}}) -\frac{1}{\sqrt{2}}\ket{V}_{3^{\prime}} (\ket{HH}_{5^{\prime}6}+\ket{VV}_{5^{\prime}6}) (\beta\ket{H}_{2^{\prime}}-\alpha\ket{V}_{2^{\prime}}).
\end{split}\label{eq:CXZ}
\end{equation}
\end{widetext}

From Eq.~\eqref{eq:CXZ}, we can obtain $\ket{\phi_{00}}$ and $\ket{\phi_{01}}$  by projecting photon 3$^{\prime}$ onto the states $\ket{+}$ and $\ket{-}$, respectively. From Eq.~\eqref{QSS:4pure-1}, we find that $\ket{\phi_{00}}(\ket{\phi_{01}})$ can be transformed into $\ket{\phi_{11}}(\ket{\phi_{10}})$ by the operation $X_{A}\otimes X_{B}\otimes I_{C}$, i.e., $X_{A}\otimes X_{B}\otimes I_{C}\ket{\phi_{00}}(\ket{\phi_{01}})=\ket{\phi_{11}}(\ket{\phi_{10}})$.
Thus, by randomly setting the degree of polarizer on photon 3$^{\prime}$ at 45$^{\circ}$($\ket{+}$) and -45$^{\circ}$($\ket{-}$) and randomly inserting two HWP set at 45$^{\circ}(X)$ on path 2$^{\prime}$ and 5$^{\prime}$, we obtain the mixed state $\rho_{\text{QSS}}$. Three photons, 2$^{\prime}$, 5$^{\prime}$ and 6 are distributed to Alice, Bob and Charlie, respectively. Note that this method used to generate $\rho_{\text{QSS}}$ is rather universal; thus, it can not only be used in linear optics systems, but also in other quantum systems as well. The detailed quantum circuit is provided in Appendix.

After preparing the quantum state for the $(3, 3)$ threshold scheme, we test the \emph{reliability} and \emph{confidentiality} on $\rho_{\text{QSS}}$. We confirm the \emph{reliability} of the $(3,3)$ threshold scheme by showing that the initial quantum information $\alpha\ket{H}+\beta\ket{V}$ issued by the dealer can be faithfully recovered by the three players. To do so, a Bell-state measurement (BSM) is first applied on Bob's and Charlie's photons. Then, conditioned on the outcome of BSM $\in\{\ket{\Phi^{+}}, \ket{\Phi^{-}}, \ket{\Psi^{+}}, \ket{\Psi^{-}}\}$, an operation $U\in\{XZ,I,Z,X\}$ is applied on Alice's photon to recover $\ket{\psi}$, where $\ket{\Phi^{\pm}}=(\ket{HH}\pm\ket{VV})/\sqrt{2}$ and $\ket{\Psi^{\pm}}=(\ket{HV}\pm\ket{VH})/\sqrt{2}$ are Bell states. As shown in Fig.~\ref{fig:setup}, the BSM is realised by interfering photons 5$^{\prime}$ and 6 on a PBS and analysed by the BSM analyser\cite{Pan98GHZa} on photon 5$^{\prime\prime}$ and photon 6$^{\prime}$. The correcting unitary $U$ is realised by HWPs. On the basis of the result from BSM analyser, we choose the corresponding unitary operation on photon 2$^{\prime}$.

In the experiment, we chose eight different input quantum states, in the form of $\ket{\psi}=\alpha\ket{H}+\beta\ket{V}$, and measure the fidelity between the input and decoded states for each case. The best fidelity achieved is 0.93$\pm$0.02 and the average fidelity is 0.82$\pm0.01$. More details on data processing are shown in Appendix.
The results are presented in Fig.~\ref{fig:recovery}. Each of the eight fidelities is beyond the classical limit 2/3 for more than 3 standard deviations.

\begin{figure}
\centering
\includegraphics[width=0.8\linewidth]{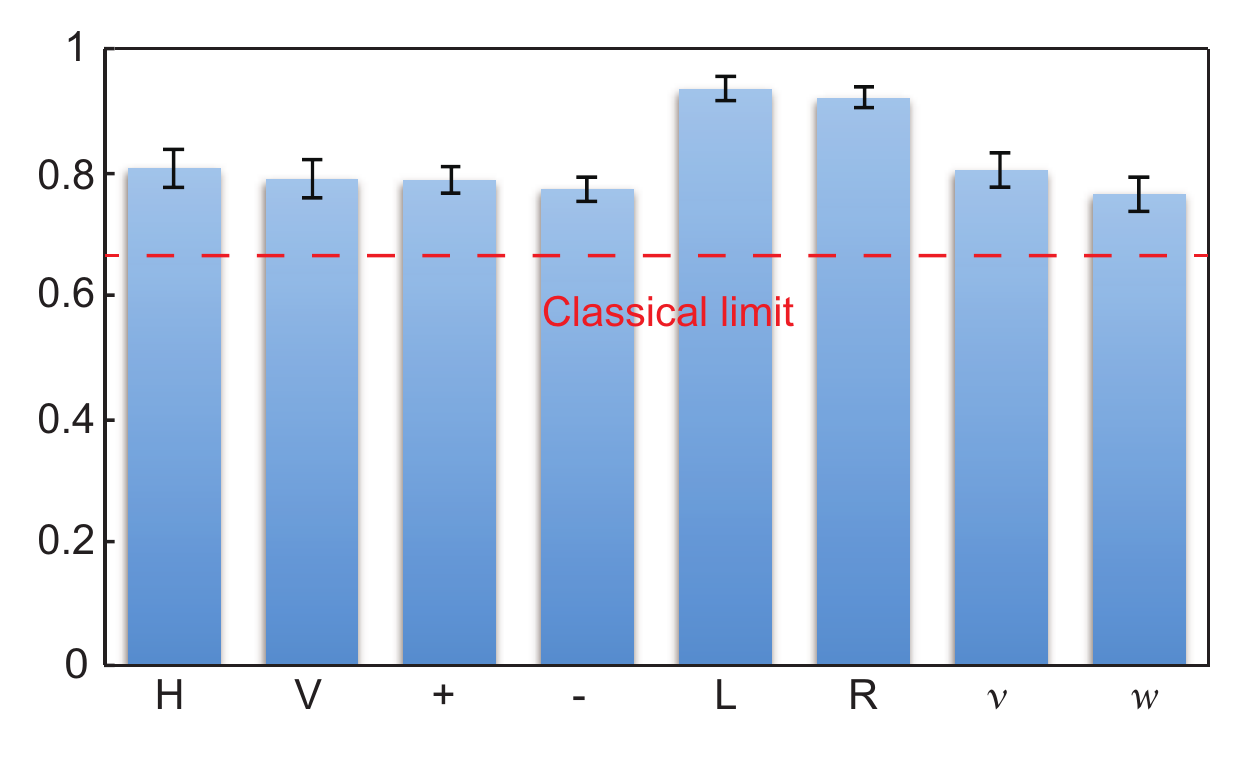}
\caption{Quantum secret recovery by three players.
The column represents the corresponding fidelity of the recovered state when the initial state is prepared into: $\ket{H(V)}$, $\ket{\pm}=(\ket{H}\pm\ket{V})/\sqrt{2}$, $\ket{L(R)}=(\ket{H}\pm i\ket{V})/\sqrt{2}$ and $\ket{v(w)}=(\ket{H}\pm\sqrt{3}\ket{V})/2$. The error bars are calculated by assuming that our experimental data follows a normal distribution. The red dashed line represents the classical limit of $2/3$.}
\label{fig:recovery}
\end{figure}

\begin{figure}[h!]
\centering
\includegraphics[width=\linewidth]{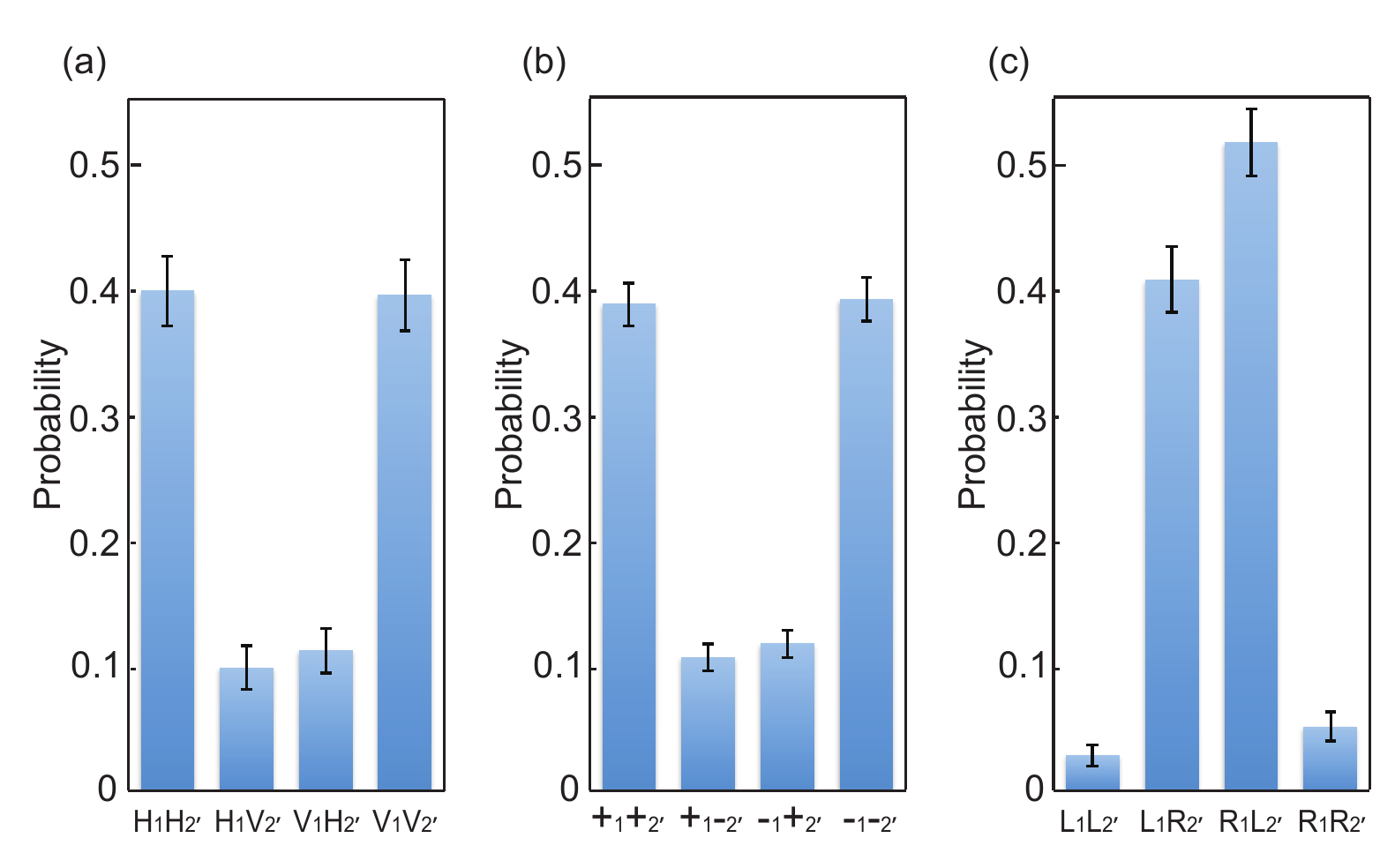}
\caption{Entanglement witness results as local measurements in the $Z$, $X$ and $Y$ bases.
(a) Coincidence detections in the $Z$ basis, projecting to $H$ and $V$,  $P(H_1,H_{2^{\prime}})=0.40(3)$, $P(H_1,V_{2^{\prime}})=0.10(2)$, $P(V_1,H_{2^{\prime}})=0.11(2)$ and $P(V_1,V_{2^{\prime}})=0.40(3)$. In Eq.~\eqref{eq:witness}, $\langle Z_{1}Z_{2^{\prime}}\rangle=P(H_{1}, H_{2^{\prime}})-P(H_{1}, V_{2^{\prime}})-P(V_{1}, H_{2^{\prime}})+P(V_{1}, V_{2^{\prime}})=0.59$.
(b) $X$ basis, projecting to $+$ and $-$, $P(+,+_{2^{\prime}})=0.40(2)$, $P(+_1,-_{2^{\prime}})=0.11(1)$, $P(-_1,+_{2^{\prime}})=0.12(1)$ and $P(-_1,-_{2^{\prime}})=0.39(2)$.
(c) $Y$ basis, projecting to $L$ and $R$, $P(L_1,L_{2^{\prime}})=0.03(1)$, $P(L_1,R_{2^{\prime}})=0.41(3)$, $P(R_1,L_{2^{\prime}})=0.52(3)$ and $P(R_1,R_{2^{\prime}})=0.05(1)$. %With the probabilities derived above, we calculate the expected value of entanglement witness as $\langle\mathcal{W}\rangle=-0.24(2)$.
}\label{fig:EPR}
\end{figure}

Furthermore, we show that entangled quantum states can also be shared and recovered by our setup. As shown in Fig.~\ref{fig:setup}, photons 1 and 2 are in a maximally entangled state $\ket{\Phi_{12}^{+}}=(\ket{HH}_{12}+\ket{VV}_{12})/\sqrt{2}$. Photon 2 is divided into three shares that are distributed to Alice, Bob and Charlie and is recovered by collaboration of all three shares. We then analyse the entanglement between photon 1 and the recovered photon 2$^{\prime}$, i.e., $\rho_{12^{\prime}}$, by an entanglement witness $\mathcal{W}=\frac{1}{2}I-\ket{\Phi_{12^{\prime}}^{+}}\bra{\Phi_{12^{\prime}}^{+}}$. The expectation value of $\mathcal{W}$ can be decomposed to a linear combination of the expectation values of local observables,
\begin{equation}\label{eq:witness}
\langle\mathcal{W}\rangle=\text{Tr}(\mathcal{W}\rho_{12^{\prime}})=\frac{1}{4}(1-\langle Z_{1}Z_{2^{\prime}}\rangle- \langle X_{1}X_{2^{\prime}}\rangle+\langle Y_{1}Y_{2^{\prime}}\rangle).
\end{equation}
The measurement results are shown in Fig.~\ref{fig:EPR}. From the measured coincidence count probabilities, we calculate that $\langle\mathcal{W}\rangle=-0.24\pm0.02$.
, from which we further obtain the fidelity of the recovered $\rho_{12^{\prime}}$ of $F_{\text{exp}}=\text{Tr}(\ket{\Phi^{+}}\bra{\Phi^{+}}\rho_{12^{\prime}})=\frac{1}{2}-\text{Tr}(\mathcal{W}\rho_{12^{\prime}})=0.74\pm0.02$\cite{Otfried07}. More details on data processing are shown in Appendix.
Clearly, the recovered quantum state is still entangled with the other half of the EPR pair because $\langle\mathcal{W}\rangle<0$. Thus, we have shown that our QSS setup is capable of sharing entangled states.

From the viewpoint of error correction, the quantum state $\rho_{\text{QSS}}$ we prepared can be treated as the 5-qubit code after going through two erasure errors. The decoding circuit shown in Fig.~\ref{fig:setup} is the same as that for the 5-qubit quantum erasure error correcting code. Thus, we have successfully demonstrated that the 5-qubit code is capable of correcting two erasure errors with a fidelity as high as 93~\%. Correcting two erasure errors has been proven to be equivalent to correcting an arbitrary error\cite{Grassl97} (details are provided in Appendix). Thus, for the first time, we have experimentally verified that a general error can be corrected in a linear optics quantum computing system.

\begin{figure*}[htbp]
\includegraphics[width=\linewidth]{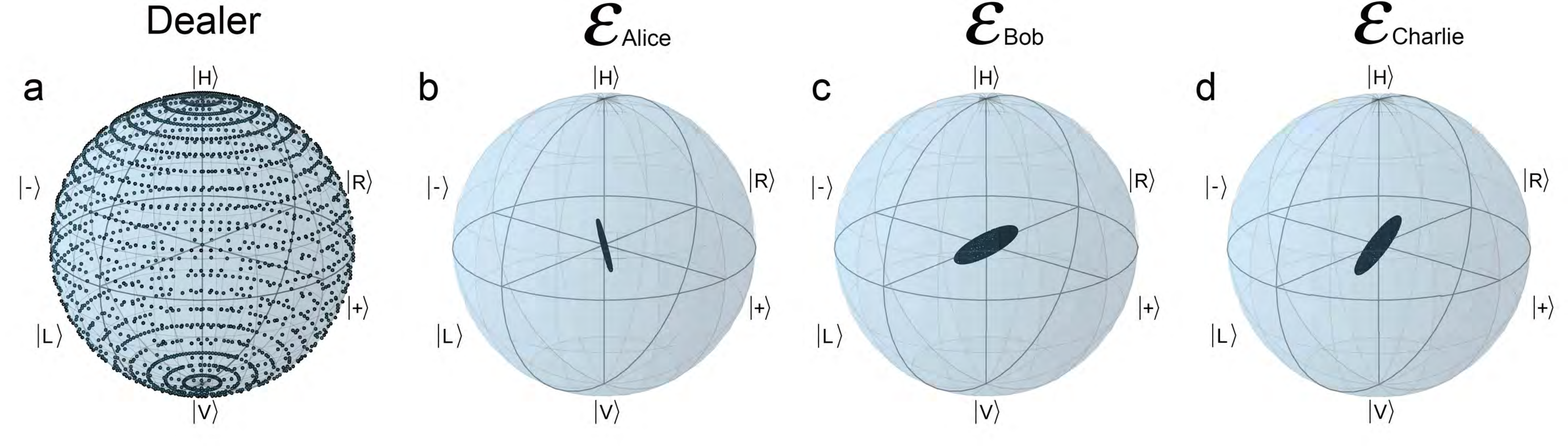}
\caption{The geometric interpretation of the channel $\mathcal{E}_{i}$ on a Bloch sphere.
Qubit states can be represented in a Bloch sphere. A pure state is on the sphere and a mixed state is in the ball. The ideal $\mathcal{E}_{ideal}$ of the quantum $(3,3)$ threshold scheme maps each point on the surface (pure state) to the origin (the maximally mixed state $I/{2}$).  (a)  Geometric interpretation of the initial pure states possibly prepared by the dealer. (b), (c), (d) Geometric interpretations of the channel $\mathcal{E}_{\text{Alice}}$, $\mathcal{E}_{\text{Bob}}$, and $\mathcal{E}_{\text{Charlie}}$, respectively.  We observe that $F_{\text{Alice}}=0.90\pm0.03$, $F_{\text{Bob}}=0.97\pm0.02$ and $F_{\text{Charlie}}=0.89\pm0.03$, where the error bars are calculated by performing 500 runs of channel operator matrix reconstructions and assuming white noise.}
\label{fig:bloch}
\end{figure*}

The \emph{confidentiality} of the scheme is shown by the fact that the quantum state of any one or two of the three players is independent of the secret quantum information $\ket{\psi}$. In an ideal implementation, from Eq.~\eqref{QSS:4pure}, we can easily find that the density matrix of each player's qubit is $I/2$ and that the density matrix of any two players' joint state is $I\otimes I/4$. Thus, no information can be obtained unless three players' shares are combined. In the experiments, we verify single-player and two-player cases separately.

In the single-player case, the encoding process can be represented by a quantum channel\cite{NC00}, i.e. $\rho_{k}=\mathcal{E}_{k}(\rho_{\text{D}})$, where $\rho_{k}$ denotes the reduced density matrix of player $k$ and $\rho_{\text{D}}=\ket{\psi}\bra{\psi}$ is the initial secret quantum state. The ideal-implementation channel $\mathcal{E}_{ideal}$ from the dealer to a single player is a depolarising channel $\rho_{D}\rightarrow I/2$.
Experimentally, we remove the PBS$_2$ and BSM analyser, and reconstruct $\mathcal{E}_{k}$  using the quantum process tomography technology\cite{NC00}. For example, when we reconstruct $\mathcal{E}_{A}$, we perform tomographic measurements on Alice while treating Bob and Charlie's qubits as trigger photons without measuring their polarisation information. The geometry interpretation of $\mathcal{E}_{k}$ is shown in Fig.~\ref{fig:bloch}. We calculate the process fidelity between ideal $\mathcal{E}_{ideal}$ and reconstructed $\mathcal{E}_{k}$, namely $F_{k}=\operatorname{Tr}\left(\sqrt{\sqrt{\mathcal{E}_{ideal}}\mathcal{E}_k\sqrt{\mathcal{E}_{ideal}}}\right)^2$ and discover that $F_{\text{Alice}}=0.90\pm0.03$, $F_{\text{Bob}}=0.97\pm0.01$ and $F_{\text{Charlie}}=0.89\pm0.03$. More details on data processing are shown in Appendix).

\begin{figure*}[htbp]
\includegraphics[width=\linewidth]{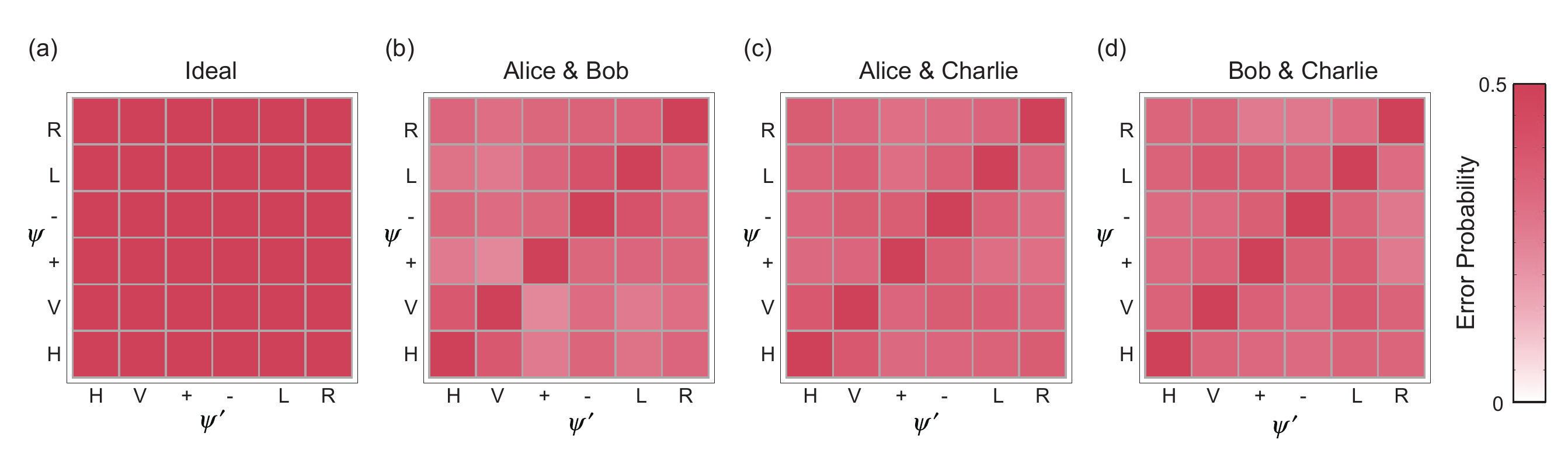}
\caption{The recovery capability of two players.
The minimum error probability $P(\varrho^{\psi}_{kl},\varrho^{\psi^{\prime}}_{kl})$ to discriminate two quantum states $\varrho^{\psi}_{kl}$ and $\varrho^{\psi^{\prime}}_{kl}$  by two players out of Alice, Bob and Charlie.
(a) $P(\varrho^{\psi}_{k,l},\varrho^{\psi^{\prime}}_{kl})$ for the ideal scenario.  (b) $P(\rho^{\psi}_{\text{AB}},\varrho^{\psi^{\prime}}_{\text{AB}})$ for Alice and Bob.  (c) $P(\varrho^{\psi}_{\text{AC}},\varrho^{\psi^{\prime}}_{\text{AC}})$ for Alice and Charlie.  (d) $P(\varrho^{\psi}_{\text{BC}},\varrho^{\psi^{\prime}}_{\text{BC}})$ for Bob and Charlie. The anti-diagonal elements represent the error probability to discriminate two identical states $P(\varrho^{\psi}_{kl},\varrho^{\psi}_{kl})$, which is always 0.5.  }
\label{fig:twoplayers}
\end{figure*}

For the two-player case, we first define the minimum error probability to distinguish two states\cite{Wilde2013}, $\varrho$ and $\varsigma$, with the optimal measurement strategy; specifically,
\begin{equation}\label{eq:MEP}
P(\varrho,\varsigma)=\frac{1}{2}(1-\frac{1}{2}||\varrho-\varsigma||),
\end{equation}
where $||\varrho-\varsigma||$ is the trace distance between $\varrho$ and $\varsigma$. $P(\varrho,\varsigma)=0$ means that $\varrho$ and $\varsigma$ can be completely distinguished via the appropriately chosen measurement basis, whereas $P(\varrho,\varsigma)=0.5$ means that $\varrho$ and $\varsigma$ are indistinguishable. We denote $\varrho_{kl}^{\psi}$ as the reduced density matrix of players $k$ and $l$ when a quantum state $\rho_{\text{D}}=\ket{\psi}\bra{\psi}$ is shared among players. For example, $\varrho_{\text{AB}}^{H}$ represents the reduced density matrix of Alice and Bob when the shared quantum state is $\ket{H}$.
For the ideal case shown in Fig.~\ref{fig:twoplayers}a, $\varrho_{kl}^{\psi}$ is $I\otimes I/4$ irrespective of what $\rho_{\text{D}}$ is, which means $P(\varrho_{kl}^{\psi},\varrho_{kl}^{\psi^{\prime}})$ is always 0.5 when $\psi$ and $\psi'$ are distinct. Experimentally, we perform the tomographic measurements on $\varrho_{\text{AB}}^{\psi}$, $\varrho_{\text{AC}}^{\psi}$ and $\varrho_{\text{BC}}^{\psi}$ for six indistinct quantum states $\rho_{\text{D}}$, and calculate the $P(\varrho_{kl}^{\psi},\varrho_{kl}^{\psi^{\prime}})$.
%Similar to the single-player case, we do the tomographic measurement on two players while treat the third one as a trigger photon with no polarization measurement on it.
As shown in Fig.~\ref{fig:twoplayers}b-d, the average $P(\varrho_{\text{AB}}^{\psi},\varrho_{\text{AB}}^{\psi^{\prime}})=0.36\pm0.01$, $P(\varrho_{\text{AC}}^{\psi},\varrho_{\text{AC}}^{\psi^{\prime}})=0.37\pm0.01$ and $P(\varrho_{\text{BC}}^{\psi},\varrho_{\text{BC}}^{\psi^{\prime}})=0.37\pm0.01$.

The following three situations are particularly interesting: $P(\varrho^{H}_{kl},\varrho^{V}_{kl})$, $P(\varrho^{+}_{kl},\varrho^{-}_{kl})$ and $P(\varrho^{L}_{kl},\varrho^{R}_{kl})$. For the dealer, $P(\ket{H}\bra{H},\ket{V}\bra{V})=P(\ket{+}\bra{+},\ket{-}\bra{-})=P(\ket{L}\bra{L},\ket{R}\bra{R})=0$, which means the dealer can perfectly distinguish \{$\ket{H},\ket{V}$\}, \{$\ket{+},\ket{-}$\} and \{$\ket{L},\ket{R}$\} by choosing an appropriate measurement basis. On the player's side, we observe the average$P(\varrho^{H}_{kl},\varrho^{V}_{kl})=0.38\pm0.02, P(\varrho^{+}_{kl},\varrho^{-}_{kl})=0.35\pm0.02$ and $P(\varrho^{L}_{kl},\varrho^{R}_{kl})=0.34\pm0.02$, which means that, even for the orthogonal input states, two players can hardly discriminate them. Therefore, we conclude that any two players cannot faithfully recover the original quantum secret state. More details on data processing are shown in Appendix.

%\section{Conclusion}
By designing a linear optical quantum circuit, we experimentally demonstrate the quantum $(3,3)$ threshold scheme, satisfying the three criteria for the fully quantum secret sharing: reliability, confidentiality, and capability of sharing entangled states. Our setup provides a practical QSS architecture. With the assistance of entanglement purification and nested entanglement swapping, a long-distance QSS scheme can be achieved to protect the secrets. Such a scheme can serve as one of the founding blocks in many quantum information tasks, such as all-photonic quantum repeater\cite{Azuma2015Allphoto,namiki2014gaussian}, distributed quantum information processing\cite{PhysRevLett.95.030505} and lossy quantum memory\cite{afzelius2010photon}.

\begin{acknowledgments}
We acknowledge H.-K.~Lo, B.~C.~Sanders and X.~Yuan for the insightful discussions. This work has been supported by the National Natural Science Foundation of China, the CAS and the National Fundamental Research Program (grant no. 2011CB921300). H.L. and Z. Z. contributed equally to this work.
\end{acknowledgments}

\pagebreak
\widetext
\begin{center}
\textbf{\large Appendix: Secret sharing of a quantum state}
\end{center}
%%%%%%%%%% Merge with supplemental materials %%%%%%%%%%
%%%%%%%%%% Prefix a "S" to all equations, figures, tables and reset the counter %%%%%%%%%%
\setcounter{equation}{0}
\setcounter{figure}{0}
\setcounter{table}{0}
\setcounter{page}{1}
\makeatletter
\renewcommand{\theequation}{S\arabic{equation}}
\renewcommand{\thefigure}{S\arabic{figure}}
\section{Theoretical description of the $(3,3)$ threshold scheme}
\subsection{Deriving the $(3,3)$ threshold form $(3,5)$ threshold}
The 5-qubit quantum error-correcting code \cite{Laflamme96}, which can be treated as a $(3,5)$ threshold scheme, encode the to-be-shared state $\alpha\ket{0}+\beta\ket{1}$ in $\alpha\ket{0_L}+\beta\ket{1_L}$, where (unnormalized)
\begin{equation} \label{QSS:33:5qubit}
\begin{aligned}
\ket{0_L}=&-\ket{00000}+\ket{01111}-\ket{10011}+\ket{11100} \\
&+\ket{00110}+\ket{01001}+\ket{10101}+\ket{11010}\\
=&-\ket{d_2}\ket{00}-\ket{d_4}\ket{11}+\ket{d_7}\ket{10}+\ket{d_5}\ket{01}, \\
\ket{1_L}=&-\ket{11111}+\ket{10000}+\ket{01100}-\ket{00011}\\
&+\ket{11001}+\ket{10110}-\ket{01010}-\ket{00101}\\
=&-\ket{d_1}\ket{11}+\ket{d_3}\ket{00}+\ket{d_8}\ket{01}-\ket{d_6}\ket{10}, \\
\end{aligned}
\end{equation}
with
\begin{equation} \label{QSS:33:ini3qubit}
\begin{aligned}
\ket{d_{_2^1}}&=(\ket{000}\pm\ket{111}), \\
\ket{d_{_4^3}}&=(\ket{100}\pm\ket{011}), \\
\ket{d_{_6^5}}&=(\ket{010}\pm\ket{101}), \\
\ket{d_{_8^7}}&=(\ket{110}\pm\ket{001}). \\
\end{aligned}
\end{equation}

In the $(3,5)$ threshold scheme, two of the five players contain no information of the to-be-shared secret.
%The 5-qubit quantum error-correcting code\cite{Laflamme96} is given by (unnormalized),
%\begin{equation} \label{QSS:33:5qubit}
%\begin{aligned}
%\ket{0_L}=&-\ket{00000}+\ket{01111}-\ket{10011}+\ket{11100} \\
%&+\ket{00110}+\ket{01001}+\ket{10101}+\ket{11010}\\
%=&-\ket{d_2}\ket{00}-\ket{d_4}\ket{11}+\ket{d_7}\ket{10}+\ket{d_5}\ket{01}, \\
%\ket{1_L}=&-\ket{11111}+\ket{10000}+\ket{01100}-\ket{00011}\\
%&+\ket{11001}+\ket{10110}-\ket{01010}-\ket{00101}\\
%=&-\ket{d_1}\ket{11}+\ket{d_3}\ket{00}+\ket{d_8}\ket{01}-\ket{d_6}\ket{10}, \\
%\end{aligned}
%\end{equation}
%where
%\begin{equation} \label{QSS:33:ini3qubit}
%\begin{aligned}
%\ket{d_{_2^1}}&=(\ket{000}\pm\ket{111}), \\
%\ket{d_{_4^3}}&=(\ket{100}\pm\ket{011}), \\
%\ket{d_{_6^5}}&=(\ket{010}\pm\ket{101}), \\
%\ket{d_{_8^7}}&=(\ket{110}\pm\ket{001}). \\
%\end{aligned}
%\end{equation}
%When the to-be-shared state is $\alpha\ket{0}+\beta\ket{1}$, a $(3,5)$ threshold can be constructed by encoding this state with $\ket{0_L}$ and $\ket{1_L}$ in Eq.~\eqref{QSS:33:5qubit}.
If two of the five players are absented, the $(3,5)$ threshold becomes to a $(3,3)$ threshold. The resultant state by discarding the two qubits in Eq.~\eqref{QSS:33:5qubit} is a three qubits mixed state $\rho_{\text{QSS}}=\frac{1}{4}\sum\limits_{i,j=0}^{1}\ket{\phi_{ij}}\bra{\phi_{ij}}$ with
%As shown in the following, the subscripts $i$ and $j$ means the measurement results of the fourth and fifth qubits.

\begin{equation}
\begin{aligned}
&\ket{00}: \phi_{00}=-\alpha\ket{d_2}+\beta\ket{d_3},\\
&\ket{11}:\phi_{11}=-\alpha\ket{d_4}-\beta\ket{d_1},\\
&\ket{10}: \phi_{10}=\alpha\ket{d_7}-\beta\ket{d_6},\\
&\ket{01}:\phi_{01}= \alpha\ket{d_5}+\beta\ket{d_8}.
\end{aligned}
\end{equation}
where the subscripts $i$ and $j$ denote the state of the two discarded qubits.

\subsection{Recovery the secret from the $(3,3)$ threshold scheme}
The three qubits of the mixed state $\rho_{\text{QSS}}$ are distributed to Alice, Bob and Charlie, respectively. In the following we describe how to recovery the secret by cooperation of Alice, Bob and Charlie.

Alice first apply a Hadamard gate on her qubit. The out-going state is,

\begin{widetext}
\begin{enumerate}
\item
$\phi_{00}$:
\begin{equation} \label{H-CNOT:00}
\begin{aligned}
-\alpha\ket{d_2}+\beta\ket{d_3}&=-\alpha(\ket{000}-\ket{111})_{\text{ABC}}+\beta(\ket{100}+\ket{011})_{\text{ABC}}\\
H_{\text{A}}&\Rightarrow-\alpha(\ket{000}+\ket{100}-\ket{011}+\ket{111})_{\text{ABC}}+\beta(\ket{000}-\ket{100}+\ket{011}+\ket{111})_{\text{ABC}}\\
&=-\alpha\ket{0}_{\text{A}}(\ket{00}-\ket{11})_{\text{BC}}-\alpha\ket{1}_{\text{A}}(\ket{00}+\ket{11})_{\text{BC}}+\beta\ket{0}_{\text{A}}(\ket{00}+\ket{11})_{\text{BC}}-\beta\ket{1}_{\text{A}}(\ket{00}-\ket{11})_{\text{BC}}\\
&=\sqrt{2}(\beta\ket{0}-\alpha\ket{1})_{\text{A}}\ket{\Phi^{+}}_{\text{BC}}-\sqrt{2}(\alpha\ket{0}+\beta\ket{1})_{\text{A}}\ket{\Phi^{-}}_{\text{BC}},\\
\end{aligned}
\end{equation}

\item
$\phi_{11}$:
\begin{equation} \label{H-CNOT:11}
\begin{aligned}
-\alpha\ket{d_4}-\beta\ket{d_1}&=-\alpha(\ket{100}-\ket{011})_{\text{ABC}}-\beta(\ket{000}+\ket{111})_{\text{ABC}}\\
H_{\text{A}}&\Rightarrow -\alpha(\ket{000}-\ket{100}-\ket{011}-\ket{111})_{\text{ABC}}-\beta(\ket{000}+\ket{100}+\ket{011}-\ket{111})_{\text{ABC}}\\
&=-\alpha\ket{0}_{\text{A}}(\ket{00}-\ket{11})_{\text{BC}}+\alpha\ket{1}_{\text{A}}(\ket{00}+\ket{11})_{\text{BC}}-\beta\ket{0}_{\text{A}}(\ket{00}+\ket{11})_{\text{BC}}-\beta\ket{1}_{\text{A}}(\ket{00}-\ket{11})_{\text{BC}}\\
&=\sqrt{2}(-\beta\ket{0}+\alpha\ket{1})_{\text{A}}\ket{\Phi^{+}}_{\text{BC}}-\sqrt{2}(\alpha\ket{0}+\beta\ket{1})_{\text{A}}\ket{\Phi^{-}}_{\text{BC}},\\
\end{aligned}
\end{equation}
\item
$\phi_{10}$:
\begin{equation} \label{H-CNOT:10}
\begin{aligned}
\alpha\ket{d_7}-\beta\ket{d_6}&=\alpha(\ket{110}+\ket{001})_{\text{ABC}}-\beta(\ket{010}-\ket{101})_{\text{ABC}}\\
H_{\text{A}}&\Rightarrow \alpha(\ket{010}-\ket{110}+\ket{001}+\ket{101})_{\text{ABC}}-\beta(\ket{010}+\ket{110}-\ket{001}+\ket{101})_{\text{ABC}}\\
&=\alpha\ket{0}_{\text{A}}(\ket{10}+\ket{01})_{\text{BC}}+\alpha\ket{1}_{\text{A}}(\ket{01}-\ket{10})_{\text{BC}}+\beta\ket{0}_{\text{A}}(\ket{01}-\ket{10})_{\text{BC}}-\beta\ket{1}_{\text{A}}(\ket{10}+\ket{01})_{\text{BC}}\\
&=\sqrt{2}(\alpha\ket{0}-\beta\ket{1})_{\text{A}}\ket{\Psi^{+}}_{\text{BC}}+\sqrt{2}(\alpha\ket{1}+\beta\ket{0})_{\text{A}}\ket{\Psi^{-}}_{\text{BC}},\\
\end{aligned}
\end{equation}

\item
$\phi_{01}$:
\begin{equation} \label{H-CNOT:01}
\begin{aligned}
\alpha\ket{d_5}+\beta\ket{d_8}&=\alpha(\ket{010}+\ket{101})_{\text{ABC}}+\beta(\ket{110}-\ket{001})_{\text{ABC}}\\
H_{\text{A}}&\Rightarrow \alpha(\ket{010}+\ket{110}+\ket{001}-\ket{101})_{\text{ABC}}+\beta(\ket{010}-\ket{110}-\ket{001}-\ket{101})_{\text{ABC}}\\
&=\alpha\ket{0}_{\text{A}}(\ket{10}+\ket{01})_{\text{BC}}-\alpha\ket{1}_{\text{A}}(\ket{01}-\ket{10})_{\text{BC}}-\beta\ket{0}_{\text{A}}(\ket{01}-\ket{10})_{\text{BC}}-\beta\ket{1}_{\text{A}}(\ket{10}+\ket{01})_{\text{BC}}\\
&=\sqrt{2}(\alpha\ket{0}-\beta\ket{1})_{\text{A}}\ket{\Psi^{+}}_{\text{BC}}-\sqrt{2}(\alpha\ket{1}+\beta\ket{0})_{\text{A}}\ket{\Psi^{-}}_{\text{BC}},\\
\end{aligned}
\end{equation}

\end{enumerate}
\end{widetext}

where we denote that four Bell states are,
 \begin{equation} \label{Bell state}
\begin{aligned}
\ket{\Phi^{+}}&=\frac{1}{\sqrt{2}}(\ket{00}+\ket{11}),\ket{\Phi^{-}}=\frac{1}{\sqrt{2}}(\ket{00}-\ket{11}),\\
\ket{\Psi^{+}}&=\frac{1}{\sqrt{2}}(\ket{01}+\ket{10}), \ket{\Psi^{-}}=\frac{1}{\sqrt{2}}(\ket{01}-\ket{10}).\\
\end{aligned}
\end{equation}

Bob and Charlie then apply a Bell state measurement (BSM) on their qubits. According to the BSM results $\{\Phi^{+},\Phi^{-},\Psi^{+},\Psi^{-}\}$, Alice can recover the initial state on her qubit by applying a unitary operation belongs to $\{XZ, I ,Z,X\}$.

\subsection{\emph{Confidentiality} of $(3,3)$ threshold scheme}

In this section, we theoretically show the \emph{confidentiality} of $(3,3)$ threshold scheme by calculating the density matrixes of all possible pairs of two-players. We find that all of the two-player reduced density matrixes are $I\otimes I/4$ regardless of what the quantum secret is.

Recall the form of $\rho_{\text{QSS}}$ after a Hadamard operation on Alice, which is also shown in Eq.\ref{H-CNOT:00}-\ref{H-CNOT:10},
\begin{widetext}
\begin{equation} \label{QSS:4puresupp}
\begin{aligned}
\rho_{\text{QSS}} &=\frac{1}{4}\sum\limits_{i,j=0}^{1}\ket{\phi_{ij}}\bra{\phi_{ij}},\\
\ket{\phi_{00}} &= \frac{1}{\sqrt{2}}[(\beta\ket{0}-\alpha\ket{1})_{\text{A}}\ket{\Psi^{+}}_{\text{BC}}-(\alpha\ket{0}+\beta\ket{1})_{\text{A}}\ket{\Psi^{-}}_{\text{BC}}],\\
\ket{\phi_{11}} &= \frac{1}{\sqrt{2}}[(-\beta\ket{0}+\alpha\ket{1})_{\text{A}}\ket{\Psi^{+}}_{\text{BC}}-(\alpha\ket{0}+\beta\ket{1})_{\text{A}}\ket{\Psi{-}}_{\text{BC}}],\\
\ket{\phi_{10}} &= \frac{1}{\sqrt{2}}[(\alpha\ket{0}-\beta\ket{1})_{\text{A}}\ket{\Phi^{+}}_{\text{BC}}+(\alpha\ket{1}+\beta\ket{0})_{\text{A}}\ket{\Phi^{-}}_{\text{BC}}],\\
\ket{\phi_{01}} &= \frac{1}{\sqrt{2}}[(\alpha\ket{0}-\beta\ket{1})_{\text{A}}\ket{\Phi^{+}}_{\text{BC}}-(\alpha\ket{1}+\beta\ket{0})_{\text{A}}\ket{\Phi^{-}}_{\text{BC}}].\\
\end{aligned}
\end{equation}
\end{widetext}

We denote $\ket{\phi}\bra{\phi}$ as $\rho(\ket{\phi})$. The subscripts $A$, $B$ and $C$ denote the quantum system of Alice, Bob and Charlie. The reduced density matrix of Alice and Bob is given by,
\begin{widetext}
\begin{equation} \label{QSS:Alice and Bob}
\begin{aligned}
\rho_{\text{AB}}=&Tr_{\text{C}}(\rho_{\text{QSS}})\\
=& \frac{1}{4}Tr_{\text{C}}[\rho_{\text{ABC}}(\ket{\phi_{00}})+\rho_{\text{ABC}}(\ket{\phi_{11}})+\rho_{\text{ABC}}(\ket{\phi_{10}})+\rho_{\text{ABC}}(\ket{\phi_{01}})]\\
=&\frac{1}{16}[\rho_{\text{AB}}(\beta\ket{00}-\alpha\ket{10}-\alpha\ket{00}-\beta\ket{10})+\rho_{\text{AB}}(\beta\ket{01}-\alpha\ket{11}+\alpha\ket{01}+\beta\ket{11})\\
&+\rho_{\text{AB}}(-\beta\ket{00}+\alpha\ket{10}-\alpha\ket{00}-\beta\ket{10})+\rho_{\text{AB}}(-\beta\ket{01}+\alpha\ket{11}+\alpha\ket{01}+\beta\ket{11})\\
&+\rho_{\text{AB}}(\alpha\ket{01}-\beta\ket{11}-\alpha\ket{11}-\beta\ket{01})+\rho_{\text{AB}}(\alpha\ket{00}-\beta\ket{10}+\alpha\ket{10}+\beta\ket{00})]\\
&+\rho_{\text{AB}}(\alpha\ket{01}-\beta\ket{11}+\alpha\ket{11}+\beta\ket{01})+\rho_{\text{AB}}(\alpha\ket{00}-\beta\ket{10}-\alpha\ket{10}-\beta\ket{00})\\
=&\frac{1}{16}\{[2|\alpha-\beta|^2+2|\alpha+\beta|^2][\rho_{\text{AB}}(\ket{00})+\rho_{\text{AB}}(\ket{11})+\rho_{\text{AB}}(\ket{01})+\rho_{\text{AB}}(\ket{10})]\}\\
=&\frac{1}{4}[\rho_{\text{AB}}(\ket{00})+\rho_{\text{AB}}(\ket{11})+\rho_{\text{AB}}(\ket{01})+\rho_{\text{AB}}(\ket{10})]\\
=&\frac{1}{4}I\otimes I.\\
\end{aligned}
\end{equation}
\end{widetext}

Considering that Bob and Charlie are symmetric, the reduced density matrix of Alice and Charlie is identical to that of Alice and Bob, which is $I\otimes I/4$.

The density matrix of Bob and Charlie is given by,
\begin{widetext}
\begin{equation} \label{QSS:Bob and Charlie}
\begin{aligned}
\rho_{\text{BC}}=&Tr_{\text{A}}(\rho_{\text{QSS}})\\
=& \frac{1}{4}Tr_{\text{A}}[\rho_{\text{ABC}}(\ket{\phi_{00}})+\rho_{\text{ABC}}(\ket{\phi_{11}})+\rho_{\text{ABC}}(\ket{\phi_{10}})+\rho_{\text{ABC}}(\ket{\phi_{01}})]\\
=&\frac{1}{16}\{\rho_{\text{BC}}[\beta(\ket{00}+\ket{11})-\alpha(\ket{00}-\ket{11})]+\rho_{\text{BC}}[-\alpha(\ket{00}+\ket{11})-\beta(\ket{00}-\ket{11})]\\
&+\rho_{\text{BC}}[-\beta(\ket{00}+\ket{11})-\alpha(\ket{00}-\ket{11})]+\rho_{\text{BC}}[\alpha(\ket{00}+\ket{11})-\beta(\ket{00}-\ket{11})]\\
&+\rho_{\text{BC}}[\alpha(\ket{10}+\ket{01})+\beta(\ket{01}-\ket{10})]+\rho_{\text{BC}}[-\beta(\ket{10}+\ket{01})+\alpha(\ket{01}-\ket{10})]\}\\
&+\rho_{\text{BC}}[\alpha(\ket{10}+\ket{01})-\beta(\ket{01}-\ket{10})]+\rho_{\text{BC}}[-\beta(\ket{10}+\ket{01})-a(\ket{01}-\ket{10})]\\
=&\frac{1}{8}[\rho_{\text{BC}}(\ket{00}+\ket{11})+\rho_{\text{BC}}(\ket{00}-\ket{11})+\rho_{\text{BC}}(\ket{10}+\ket{01})+\rho_{\text{BC}}(\ket{10})-\ket{01}]\\
=&\frac{1}{4}[\rho_{\text{BC}}(\ket{00})+\rho_{\text{BC}}(\ket{11})+\rho_{\text{BC}}(\ket{01})+\rho_{\text{BC}}(\ket{10})]\\
=&\frac{1}{4}I\otimes I.\\
\end{aligned}
\end{equation}
\end{widetext}

According to above results, all the density matrixes of two players are $I\otimes I/4$. Hence, two players know no information about the initial state. It is easy to see that the density matrix of a single player must be $I/2$. Then we prove the confidentiality of our scheme.

\subsection{Quantum circuit for the $(3, 3)$ threshold scheme}

\begin{figure*}
\centering
\includegraphics[width=\linewidth]{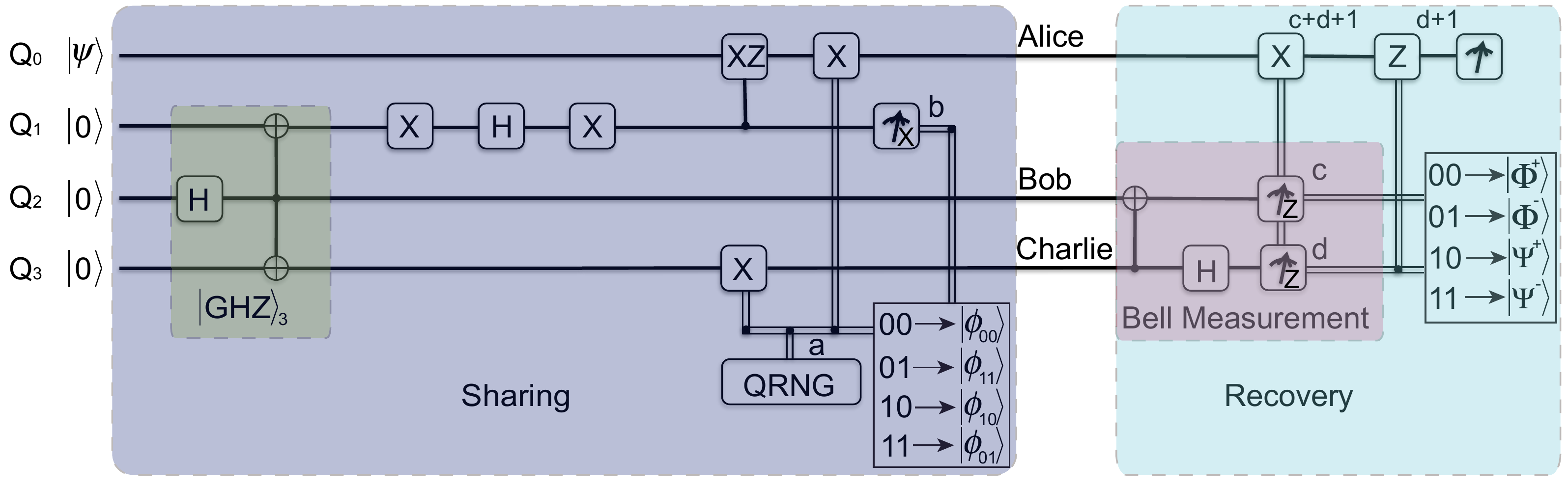}
\caption{\textbf{Schematic diagram for the circuit of the QSS scheme.}
The dealer uses a QRNG to generate a random bit $a\in \{0, 1\}$, according to which an $X$ operation is applied to $Q_0$ and $Q_3$. After certain operations, the dealer performs $X$ basis measure on $Q_1$ to obtain a classical bit $b\in\{0, 1\}$, which should be uniformly random if the state preparation is perfect. According to two bits $ab\in\{00,01,10,11\}$, the dealer knows that the out-going state corresponds to $\ket{\phi}\in$ \{$\ket{\phi_{00}}, \ket{\phi_{11}}, \ket{\phi_{10}}, \ket{\phi_{01}}$\}, respectively. Therefore, the prepared state can be treated as a three-qubit mixed state $\rho_{\text{QSS}}$ in Eq.~\ref{QSS:4puresupp}. On the recovery stage, the operation $X^{c+d+1}Z^{d+1}$ is applied on Alice to reconstruct the initial state $\ket{\psi}$ according to the BSM results $cd\in\{00, 01, 10, 11\}$ on Bob and Charlie.}
\label{fig:circuit}
\end{figure*}

A quantum circuit design for the $(3,3)$ threshold scheme is presented in Fig.~\ref{fig:circuit}. We start the scheme with four qubits. A dealer possesses a quantum state $\ket{\psi}=\alpha\ket{0}+\beta\ket{1}$, and entangles it with a three-qubit GHZ state $\ket{\text{GHZ}}_{3}=(\ket{000}+\ket{111})/\sqrt{2}$, which is composed by $Q_1$, $Q_2$ and $Q_3$, by a two-qubit  entangling gate controlled-$XZ$ between $Q_{0}$ and $Q_1$. Here, we denote $X$, $Y$, and $Z$ to be Pauli operations. Two $X$ operations, which are conditioned on a classical random bit $a\in\{0,1\}$ generated by a quantum random number generator (QRNG), are applied on $Q_0$ and $Q_3$. $Q_1$ is randomly projected to the eigenstate of the $X$ basis, where the result leads another classical random bit $b\in\{0,1\}$. Then the out-going three-qubit state $\ket{\phi_{ij}}\in\{\ket{\phi_{00}}, \ket{\phi_{11}}, \ket{\phi_{10}}, \ket{\phi_{01}}\}$ is distributed to Alice, Bob and Charlie, each of who will get one qubit. Here, $\ket{\phi_{ij}}$ is conditioned on the two classical bits $a$ and $b$. In our scheme, $a, b$ are chosen from 00, 01, 10 and 11 with equal probabilities $25\%$, which leads the mixed state $\rho_{\text{QSS}}$, given in Eq.~\ref{QSS:4puresupp}.
\subsection{$t$ bits arbitrary errors $=$ $2t$ bits erasure errors }
In this section, we will prove that a code can correct $t$ arbitrary errors at unknown positions is equivalent to a $2t$ erasure error-correcting code.

We know that an arbitrary $2^n\times 2^n$ matrix that acts on an n qubits space can be considered in terms of Pauli operators $\{I, X, Y, Z\}^{\otimes n}$. Then, we denote $\epsilon\in \{I, X, Y, Z\}^{\otimes n}$ as the error subset contains all errors we wish to correct.
According to\cite{PhysRevA.55.900}, we know that the necessary and sufficient condition on quantum error-correcting code can correct all errors in $\epsilon$ is,
 \begin{equation} \label{correctable}
\begin{aligned}
\bra{j}E_b^\dag E_a\ket{i}=C_{ba}\delta_{ij},
\end{aligned}
\end{equation}
where $E_a,E_b \in \epsilon$ and  $\{\ket{i}\}$ is an orthonormal basis of the code subspace. $C_{ab}$ is an Hermitian matrix which is independent with $\ket{i}$ and $\ket{j}$. We denote that $\delta_{ij}=0$ if $i\ne j$, and $\delta_{ij}=0$, otherwise. We denote that the weight of a $E_a\in \epsilon$ is number of nontrivial Pauli operators acting on qubits.

When considering the $t$-arbitrary error-correcting code, the weights of $E_b^\dag$ and $E_a$ are at most $t$ and therefore, the set of $\epsilon_1=\{E_b^\dag E_a\}$ is all Pauli operators with weight no more than $2t$.

Correcting erasure error is equivalent to correct arbitrary located error. Firstly, erasure error is a special case of the arbitrary located error. Secondly, if the location of a error is known, we can directly this qubit and the error changes to erasure error. Hence correcting erasure error is equivalent to correct arbitrary located error.

When correct $2t$ erasure errors, the weights of $E_b^\dag$ and $E_a$ are $2t$. We know that two pauli operators act on one qubits is still a pauli operator.
Since the errors act on $2t$ specified location and therefore, the weight of $E_b^\dag E_a$ is $2t$. Then the set $\epsilon_2$ of all $E_b^\dag E_a$ only contains all Pauli operators with weight no more than $2t$.

Then we know that $\epsilon_1=\epsilon_2$. According to Eq.~\eqref{correctable}, we know that a $t$-arbitrary error-correcting code is a $2t$ erasure error-correcting code\cite{PhysRevA.56.33}.

\section{Further experimental details}

\subsection{Photon source}
In our experiments, a femtosecond pulse (130fs, 76MHz, 780nm, 3.4W) is converted into an ultraviolet pulse (390nm, 1.4W) through a frequency doubling $\text{LiB}_3\text{O}_5$ crystal. The ultraviolet pulse then passes through three BBO crystals and creates three entangled photon pairs via spontaneous parametric down-conversion (SPDC).

To suppress the high-order emission in SPDC, we attenuate the power of ultraviolet pulses to 800mW, which is an appropriate power for a high counter rate and high visibility.
For the state preparation, we set narrow band-pass filters with a full-width at half of the transmittance maximum (FWHM) of $\Delta\lambda_{\text{FWHM}}=8$~nm on ordinary photons (1, 4$^{\prime}$ and 5$^{\prime}$) and  narrow band-pass filters with a FWHM of $\Delta\lambda_{\text{FWHM}}=3$~nm on extraordinary photons (2$^{\prime}$, 3$^{\prime}$ and 6).
With this filter setting, we observe the two-folder coincidences for $\ket{\Phi_{12^{\prime}}^{+}}$, $\ket{\Phi_{3^{\prime}4^{\prime}}^{+}}$ and $\ket{\Phi_{5^{\prime}6}^{+}}$ to be 1.74$\times 10^{5}$, 1.26$\times10^{5}$ and 1.06$\times10^{5}$ counts per second (cps), respectively, with an average overall detection efficiency of 23\% and a generation rate of 0.033 per pulse. The Hong-Ou-Mandel type visibilities of the interference on the PDBS and PBS$_1$ are 53.4\% and 69.2\%, respectively. Finally, we observer the the sixfold coincidence count rate of 75 cph.

When we proceed to the stage of recovering the quantum secret, we need apply a BSM on photons 5$^{\prime}$ and 6. Note that photons 5$^{\prime}$ and 5 are o-ray and e-ray, respectively. To achieve good indistinguishability on PBS$_{2}$, we replace the narrow band-pass filter $\Delta\lambda_{\text{FWHM}}=8$~nm on paths 5$^{\prime}$ and 4$^{\prime}$ with narrow band-pass filters $\Delta\lambda_{\text{FWHM}}=3$~nm on paths 5$^{\prime\prime}$ and 4$^{\prime}$, respectively. With this filter setting, the sixfold coincidence count rate is reduced to 9 cph.

\subsection{Decomposition of controlled-$XZ$ operation}
We show that the controlled-XZ gate can be decomposed to a controlled-NOT(CNOT) gate and three phase shift gates.

\begin{figure}[h!]
\centering
\includegraphics[width=\linewidth]{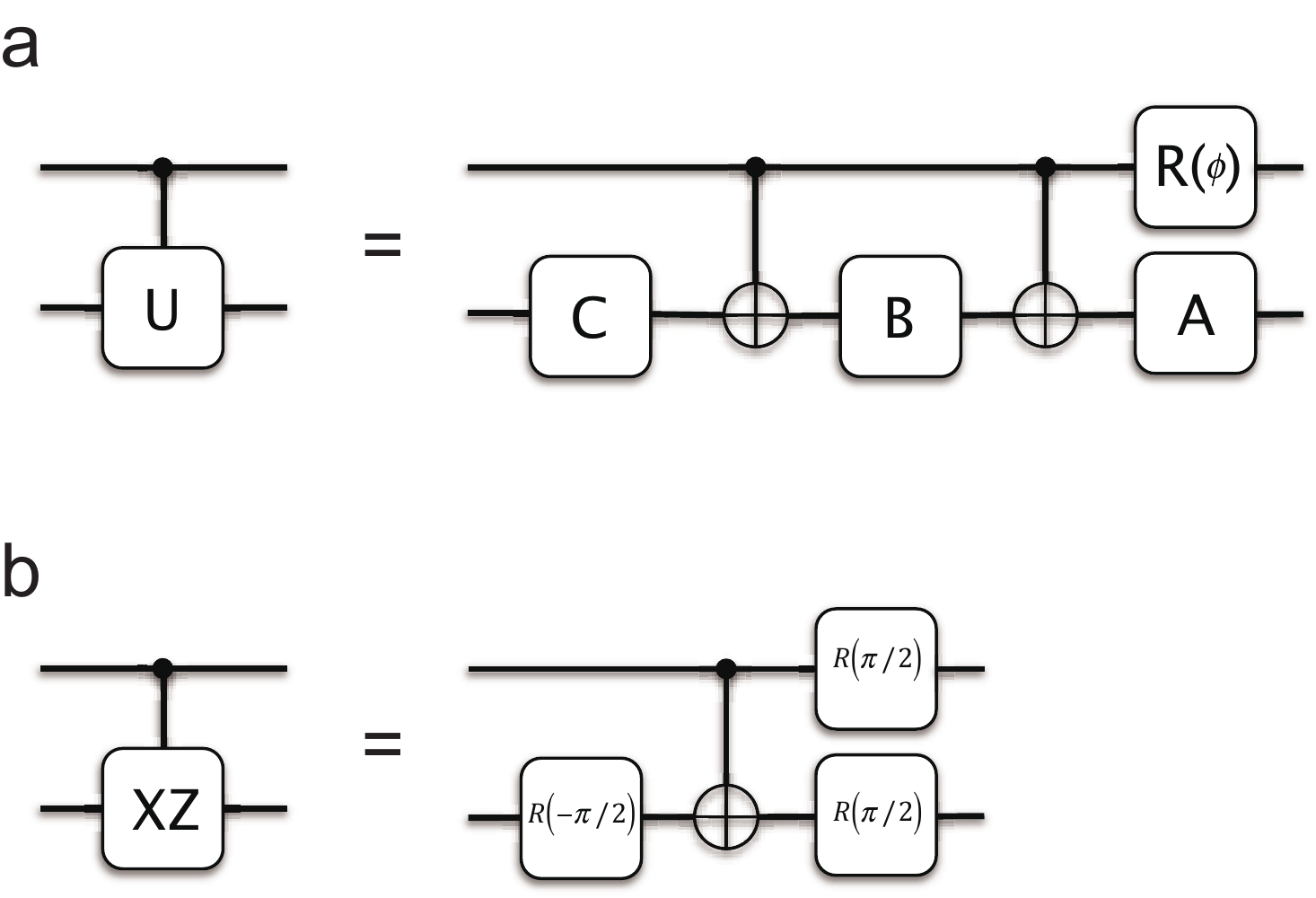}
\caption{\textbf{Decomposition of controlled-$U$ operation .}
\textbf{a,} The decomposition of general controlled-$U$ operation.
\textbf{b,} The decomposition of controlled-XZ operation.
\label{decomposition}}
\end{figure}

The generic quantum circuit to implement the controlled-U operation is shown in Fig.~\ref{decomposition}a. $U$ is in the form $U=e^{i\phi}AXBXC$, where $A$, $B$ and $C$ are single-qubit operations and satisfy $ABC = I$. $R(\phi)$ is the single qubit phase shift gate,
\begin{equation}
R(\phi)=
\begin{pmatrix}
1 & 0\\
0 & e^{i\phi}\\
\end{pmatrix}
\end{equation}
When the control qubit is set at 0, $ABC=I$ act on the target qubit. When the control qubit is set at 1, $U=e^{i\phi} AXBXC$ act on the target qubit \cite{NC00}.

In our case, the unitary is
\begin{equation}
U=X\cdot Z=
\begin{pmatrix}
0 & -1\\
1 & 0\\
\end{pmatrix}
\end{equation}
We find that $U$ can be represent by $U=e^{i\pi/2}\cdot R(\pi/2) \cdot X\cdot R(-\pi/2)$, with $R(\pi/2)\cdot R(-\pi/2)=I $.
The quantum circuit to decompose controlled-XZ operation is shown in Fig.~\ref{decomposition}b.

\subsection{Secret recovery with correcting operation}

In this section, we show more details of the recovery of the quantum secret, i.e. \emph{reliability} of the $(3, 3)$ scheme.

\subsubsection{Recovery of the single-photon state}

\begin{figure}[b!]
\centering
\includegraphics[width=\linewidth]{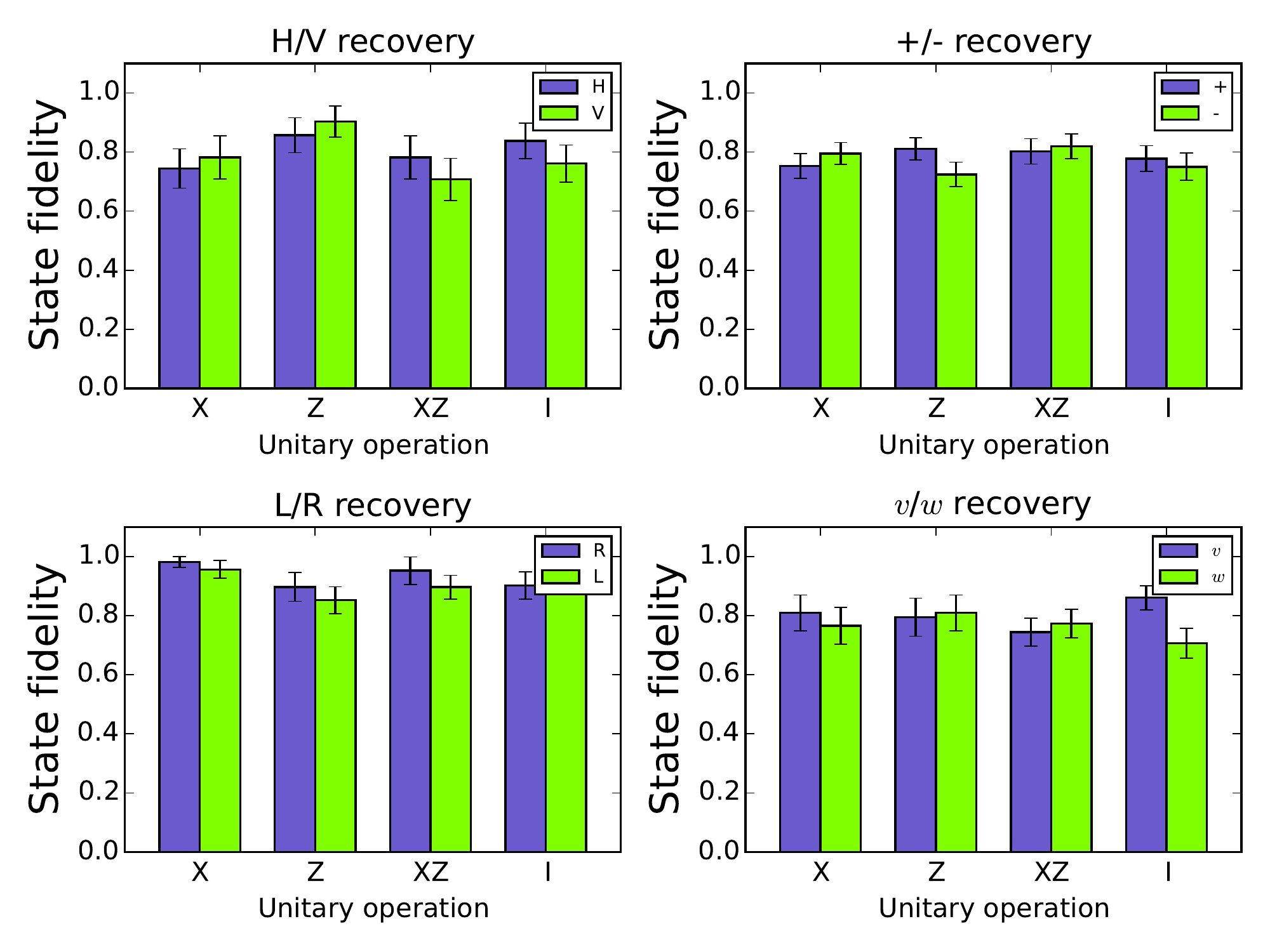}
\caption{\textbf{Secret recovery of eight distinct states with according unitary operation.} In \textbf{a, b, c, } and \text{d,} the dealer respectively prepares the initial state in $\ket{H(V)}$, $\ket{\pm}$, $\ket{L(R)}$ and $ \ket{v(w)}$. According to results of BSM result $\in\{\ket{\Phi^{+}}, \ket{\Phi^{-}}, \ket{\Psi^{+}}, \ket{\Psi^{-}}\}$, an unitary operation performs on the state of Alice to recover the state. For each possible scenario, we calculate the fidelity of the recovering state and its error bar.
\label{fig:recoverysupp}}
\end{figure}

As discussed in the main text, conditioned on the result of BSM result $\in\{\ket{\Phi^{+}}, \ket{\Phi^{-}}, \ket{\Psi^{+}}, \ket{\Psi^{-}}\}$, a unitary operation $\in\{ ZX, I, Z, X\}$ is applied on Alice to correct to state. In the experiment, the unitary operation can be realized by using two HWPs, the $X$ operation is a HWP setting at 45$^{\circ}$ and the $Z$ operation is a HWP setting at 0$^{\circ}$. As shown in the Fig.\ref{fig:recoverysupp}, we assume that the dealer prepares the state in $\ket{H(V)}=\ket{0(1)}$, $\ket{\pm}=(\ket{0}\pm\ket{1})/\sqrt{2}$, $\ket{L(R)}=\ket{0}\pm i\ket{1}$ and $ \ket{v(w)}=(\ket{0}\pm\sqrt{3}\ket{1})/2$, respectively. Fidelity of the recovering state is calculated in each case when the dealer prepared the initial state in one of above eight states and performs one of the four possible unitary operations to recover this state.

\subsubsection{Recovery of two-photon entanglement}

When we analyze the entanglement between photon 1 and the recovered photon 2$^{\prime}$, we measure the entanglement witness $\mathcal{W}=\frac{1}{2}I-\ket{\Phi_{12^{\prime}}^{+}}\bra{\Phi_{12^{\prime}}^{+}}$ between these two photons, where $\ket{\Phi_{12}^{+}}=(\ket{HH}_{12}+\ket{VV}_{12})/\sqrt{2}$. If the state of two photons is a separable state, the value of the witness should be $0$. If the value of the entanglement witness is smaller than $0$, two photons are entangled with each other. As shown in Eq.~4 in the main text, the expectation value of the entanglement witness can be written as

\begin{equation}\label{eq:witnesssupp}
\langle\mathcal{W}\rangle=\frac{1}{4}(1-\langle Z_{1}Z_{2^{\prime}}\rangle- \langle X_{1}X_{2^{\prime}}\rangle+\langle Y_{1}Y_{2^{\prime}}\rangle).
\end{equation}

Experimentally, to get the expectation value of $\langle Z_{1}Z_{2^{\prime}}\rangle$, $\langle X_{1}X_{2^{\prime}}\rangle$ and $\langle Y_{1}Y_{2^{\prime}}\rangle$, we project photon 1 and 2$^{\prime}$ on the eigenstate of $Z$, $X$ and $Y$ as the other photons are measured according the recovery procedure described in the main text. The probabilities shown in Fig.~3 is calculated by
\begin{widetext}
\begin{equation}\label{eq:probability}
\begin{split}
&P(H_1, H_{2^{\prime}})=\frac{N_{H_1, H_{2^{\prime}}}}{N_{H_1, H_{2^{\prime}}}+N_{H_1, V_{2^{\prime}}}+N_{V_1, H_{2^{\prime}}}+N_{V_1, V_{2^{\prime}}}},
P(H_1, V_{2^{\prime}})=\frac{N_{H_1, V_{2^{\prime}}}}{N_{H_1, H_{2^{\prime}}}+N_{H_1, V_{2^{\prime}}}+N_{V_1, H_{2^{\prime}}}+N_{V_1, V_{2^{\prime}}}},\\
&P(V_1, H_{2^{\prime}})=\frac{N_{V_1, H_{2^{\prime}}}}{N_{H_1, H_{2^{\prime}}}+N_{H_1, V_{2^{\prime}}}+N_{V_1, H_{2^{\prime}}}+N_{V_1, V_{2^{\prime}}}},
P(V_1, V_{2^{\prime}})=\frac{N_{V_1, V_{2^{\prime}}}}{N_{H_1, H_{2^{\prime}}}+N_{H_1, V_{2^{\prime}}}+N_{V_1, H_{2^{\prime}}}+N_{V_1, V_{2^{\prime}}}},\\
&P(+_1, +_{2^{\prime}})=\frac{N_{+_1, +_{2^{\prime}}}}{N_{+_1, +_{2^{\prime}}}+N_{+_1, -_{2^{\prime}}}+N_{-_1, +_{2^{\prime}}}+N_{-_1, -_{2^{\prime}}}},
P(+_1, -_{2^{\prime}})=\frac{N_{+_1, -_{2^{\prime}}}}{N_{+_1, +_{2^{\prime}}}+N_{+_1, -_{2^{\prime}}}+N_{-_1, +_{2^{\prime}}}+N_{-_1, -_{2^{\prime}}}},\\
&P(-_1, +_{2^{\prime}})=\frac{N_{-_1, +_{2^{\prime}}}}{N_{+_1, +_{2^{\prime}}}+N_{+_1, -_{2^{\prime}}}+N_{-_1, +_{2^{\prime}}}+N_{-_1, -_{2^{\prime}}}},
P(-_1, -_{2^{\prime}})=\frac{N_{+_1, +_{2^{\prime}}}}{N_{+_1, +_{2^{\prime}}}+N_{+_1, -_{2^{\prime}}}+N_{-_1, +_{2^{\prime}}}+N_{-_1, -_{2^{\prime}}}},\\
&P(L_1, L_{2^{\prime}})=\frac{N_{L_1, L_{2^{\prime}}}}{N_{L_1, L_{2^{\prime}}}+N_{L_1, R_{2^{\prime}}}+N_{R_1, L_{2^{\prime}}}+N_{R_1, R_{2^{\prime}}}},
P(L_1, R_{2^{\prime}})=\frac{N_{L_1, R_{2^{\prime}}}}{N_{L_1, L_{2^{\prime}}}+N_{L_1, R_{2^{\prime}}}+N_{R_1, L_{2^{\prime}}}+N_{R_1, R_{2^{\prime}}}},\\
&P(R_1, L_{2^{\prime}})=\frac{N_{R_1, L_{2^{\prime}}}}{N_{L_1, L_{2^{\prime}}}+N_{L_1, R_{2^{\prime}}}+N_{R_1, L_{2^{\prime}}}+N_{R_1, R_{2^{\prime}}}},
P(R_1, R_{2^{\prime}})=\frac{N_{R_1, R_{2^{\prime}}}}{N_{L_1, L_{2^{\prime}}}+N_{L_1, R_{2^{\prime}}}+N_{R_1, L_{2^{\prime}}}+N_{R_1, R_{2^{\prime}}}},\\
\end{split}
\end{equation}
\end{widetext}
where the $N$ is the coincidence count, e.g., $N_{H_1, H_{2^{\prime}}}$ is the number of events when photon 1 and 2$^{\prime}$ are projected on $\ket{H}$ and $\ket{H}$, respectively. With the probabilities in Eq.~\ref{eq:probability}, we can calculate $\langle Z_{1}Z_{2^{\prime}}\rangle$, $\langle X_{1}X_{2^{\prime}}\rangle$ and $\langle Y_{1}Y_{2^{\prime}}\rangle$ by
\begin{equation}\label{eq:expectation}
\begin{split}
&\langle Z_{1}Z_{2^{\prime}}\rangle\\
&=P(H_1, H_{2^{\prime}})-P(H_1, V_{2^{\prime}})-P(V_1, H_{2^{\prime}})+P(V_1, V_{2^{\prime}}),\\
&\langle X_{1}X_{2^{\prime}}\rangle\\
&=P(+_1, +_{2^{\prime}})-P(+_1, -_{2^{\prime}})-P(-_1, +_{2^{\prime}})+P(-_1, -_{2^{\prime}}),\\
&\langle Y_{1}Y_{2^{\prime}}\rangle\\
&=P(L_1, L_{2^{\prime}})-P(L_1, R_{2^{\prime}})-P(R_1, L_{2^{\prime}})+P(R_1, R_{2^{\prime}}).\\
\end{split}
\end{equation}

The experimental results of probabilities in Eq.~\ref{eq:probability}n and expectation values in Eq.~\ref{eq:expectation} are in TABLE.~\ref{probability}. The numbers in the parentheses is the calculated errors by assuming that the raw count statistic is Poissonian, i.e., $P(H_1, H_{2^{\prime}})$ is equal to $0.40\pm 0.03$. Finally, the value of $\langle\mathcal{W}\rangle$ is equal to $-0.24\pm0.02$, which means that the photon 1 and 2$^{\prime}$ are entangled with each other.
\begin{table}[h!]
\centering
 \begin{tabular}{|c|c|c|c|c|c|}
 \hline
 $P(H_1, H_{2^{\prime}})$ & 0.40(3) & $P(+_1, +_{2^{\prime}})$&0.40(2) & $P(L_1, L_{2^{\prime}})$ & 0.03(1)\\
 \hline
$P(H_1, V_{2^{\prime}})$ & 0.10(2) & $P(+_1, -_{2^{\prime}})$& 0.11(1)& $P(L_1, R_{2^{\prime}})$ & 0.41(3) \\
\hline
$P(V_1, H_{2^{\prime}})$ & 0.28(2)& $P(-_1, +_{2^{\prime}})$& 0.12(1) & $P(R_1, L_{2^{\prime}})$ & 0.52(3)\\
\hline
$P(V_1, V_{2^{\prime}})$& 0.11(2)& $P(-_1, -_{2^{\prime}})$& 0.39(2)& $P(R_1, R_{2^{\prime}})$& 0.05(1)\\
\hline
$\langle Z_{1}Z_{2^{\prime}}\rangle$& 0.58(5) & $\langle X_{1}X_{2^{\prime}}\rangle$ &0.56(3) &$\langle Y_{1}Y_{2^{\prime}}\rangle$ & -0.84(3)\\
\hline
 \end{tabular}
 \caption{\textbf{Coincidence probabilities and expectation value of $Z_{1}Z_{2^{\prime}}$, $X_{1}X_{2^{\prime}}$ and $Y_{1}Y_{2^{\prime}}$.}  }
 \label{probability}
 \end{table}

\subsection{Quantum process tomography between d    ealer and single player}

\begin{figure}
\centering
\includegraphics[width=0.8\linewidth]{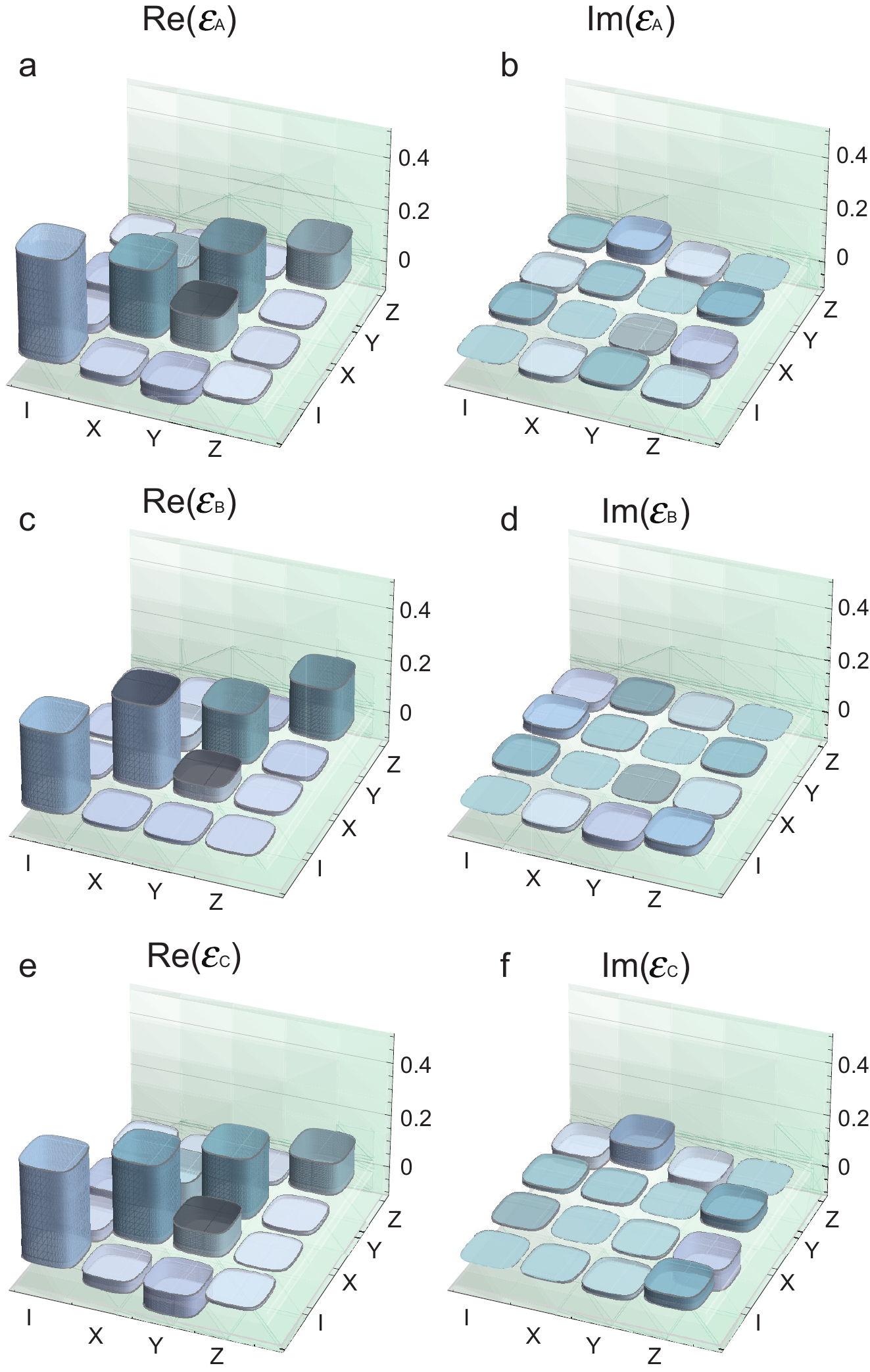}
\caption{\textbf{Reconstructed density matrix of $\mathcal{E}_{k}$ .}
\textbf{a, b,} Real and imaginary part of $\mathcal{E}_{\text{A}}$.
\textbf{c, d,} Real and imaginary part of $\mathcal{E}_{\text{B}}$.
\textbf{e, f,} Real and imaginary part of $\mathcal{E}_{\text{C}}$.
\label{processmatrix}}
\end{figure}

In this section, we show the reconstructed process density matrix of $\mathcal{E}_{k}$.
We use the quantum process tomography (QPT) technology to reconstruct the process matrix $\mathcal{E}_{k}$ ($\mathcal{E}_{\text{A}}$, $\mathcal{E}_{\text{B}}$ or $\mathcal{E}_{\text{C}}$) between the dealer and the single player (Alice, Bob or Charlie). An ideal $\mathcal{E}_k$ is a depolarizing channel $\rho\to(1-\lambda)\rho+\frac{\lambda}{3}(X\rho X+Y\rho Y+Z\rho Z)$ with $\lambda=3/4$. Hence, according to the reconstructed density matrix shown in Fig.~\ref{processmatrix}, we can respectively calculate process fidelities $F_\text{Alice} = 0.90 \pm 0.03$, $F_\text{Bob} = 0.97 \pm 0.01$ and $F_\text{Charlie} = 0.89 \pm 0.03$.

\subsection{Minimum error probability in the \emph{confidentiality} test of two players}

The minimum probabilities in the \emph{confidentiality} test of Alice \& Bob, Alice \& Charlie and Bob \& Charlie are shown in TABLE~\ref{AandB}, TABLE~\ref{AandC} and TABLE~\ref{BandC}, respectively. Here the numbers of absolute errors ($0.01$) are list in the parentheses. For example, as shown in TABLE~\ref{AandB}, the probability $P(\varrho_{\text{AB}}^{L},\varrho_{\text{AB}}^{H})=0.34\pm 0.02$.
\begin{table}[h!]
\centering
 \begin{tabular}{|c|c|c|c|c|c|c|}
 \hline
& $\ket{H}$ & $\ket{V}$ & $\ket{+}$ &$\ket{-}$ & $\ket{R}$ & $\ket{L}$\\
 \hline
 $\ket{L}$
& 0.34(2) & 0.31(1) & 0.33(1)& 0.35(2)& 0.35(1) & 0.5 \\
$\ket{R}$
& 0.30(1) & 0.28(2)& 0.35(1)& 0.41(2) & 0.5 & 0.35(1)\\
$\ket{-}$
& 0.34(2)& 0.32(2)& 0.33(1)& 0.5& 0.41(2)& 0.35(2)\\
$\ket{+}$
& 0.27(2)& 0.23(1)& 0.5& 0.33(1)& 0.35(1)& 0.33(1)\\
$\ket{V}$
& 0.39(1)& 0.5& 0.23(1)& 0.32(1)& 0.28(2)& 0.31(1)\\
$\ket{H}$
&0.5& 0.39(1)& 0.27(2)& 0.34(2)& 0.30(1)& 0.34(2)\\
\hline
 \end{tabular}
 \caption{\textbf{$P(\varrho_{\text{AB}}^{\psi},\varrho_{\text{AB}}^{\psi^{\prime}})$ of Alice and Bob.} The left-most column are the six input states of $\ket{\psi}$ and the top row are the six input states of $\ket{\psi^{\prime}}$. }
 \label{AandB}
 \end{table}

\begin{table}[h!]
\centering
 \begin{tabular}{|c|c|c|c|c|c|c|}
 \hline
 & $\ket{H}$ & $\ket{V}$ & $\ket{+}$ &$\ket{-}$ & $\ket{R}$ & $\ket{L}$\\
 \hline
 $\ket{L}$
& 0.38(2)& 0.34(1)& 0.30(1)& 0.32(1)& 0.35(2)& 0.5 \\
$\ket{R}$
& 0.35(1)& 0.37(1)& 0.31(1)& 0.36(1)& 0.5& 0.35(2)\\
$\ket{-}$
& 0.34(2)& 0.38(2)& 0.37(2)& 0.5& 0.36(1)& 0.32(1)\\
$\ket{+}$
& 0.33(1)& 0.34(1)& 0.5& 0.37(2)& 0.31(1)& 0.30(1)\\
$\ket{V}$
& 0.39(2)& 0.5& 0.34(1)& 0.38(2)& 0.37(1)& 0.34(1)\\
$\ket{H}$
& 0.5& 0.39(2)& 0.33(1)& 0.34(2)& 0.35(1)& 0.38(2)\\
\hline
 \end{tabular}
 \caption{\textbf{$P(\varrho_{\text{AC}}^{\psi},\varrho_{\text{AC}}^{\psi^{\prime}})$ of Alice and Charlie.} The left-most column are the six input states of $\ket{\psi}$ and the top row are the six input states of $\ket{\psi^{\prime}}$.}
 \label{AandC}
 \end{table}

\begin{table}[!htb]
 \begin{tabular}{|c|c|c|c|c|c|c|}
 \hline
 & $\ket{H}$ & $\ket{V}$ & $\ket{+}$ &$\ket{-}$ & $\ket{R}$ & $\ket{L}$\\
 \hline
$ \ket{L}$
& 0.34(1)& 0.35(2)& 0.28(1)& 0.28(1)& 0.32(1)& 0.5 \\
$\ket{R}$
& 0.35(2)& 0.40(2)& 0.38(2)& 0.35(2)& 0.5& 0.32(1)\\
$\ket{-}$
&0.32(1)& 0.33(2)& 0.36(2)& 0.5& 0.35(2)& 0.28(1)\\
$\ket{+}$
&0.33(1)& 0.36(2)& 0.5& 0.36(2)& 0.38(2)& 0.28(1)\\
$\ket{V}$
& 0.35(2)& 0.5& 0.36(2)& 0.33(2)& 0.40(2)& 0.35(2)\\
$\ket{H}$
& 0.5& 0.35(2)& 0.33(1)& 0.32(1)& 0.35(2)& 0.34(1)\\
\hline
\end{tabular}
\caption{\textbf{$P(\varrho_{\text{BC}}^{\psi},\varrho_{\text{BC}}^{\psi^{\prime}})$ of Bob and Charlie.} The left-most column are the six input states of $\ket{\psi}$ and the top row are the six input states of $\ket{\psi^{\prime}}$.}
 \label{BandC}
 \end{table}
 
 \bibliography{bibQSS}

\end{document}